\documentclass[aps,prb,twocolumn,superscriptaddress,preprintnumbers,amsmath,amssymb,floatfix]{revtex4}

\usepackage{graphicx}
\usepackage{longtable}
\usepackage{bm}
\usepackage{multirow}
\usepackage{hyperref}
\usepackage{amssymb}
\usepackage{amsmath}
\usepackage{array}

\begin{document}

\title{First-order structural transition and pressure-induced lattice/phonon anomalies in Sr$_2$IrO$_4$}

\author{K. Samanta}
\affiliation{``Gleb Wataghin'' Institute of Physics, University of Campinas - UNICAMP, Campinas, S\~ao Paulo 13083-859, Brazil}

\author{F. M. Ardito}
\affiliation{Brazilian Synchrotron Light Laboratory (LNLS), Brazilian Center for Research in Energy and Materials (CNPEM), Campinas, S\~ao Paulo 13083-970, Brazil}

\author{N. M. Souza-Neto}
\affiliation{Brazilian Synchrotron Light Laboratory (LNLS), Brazilian Center for Research in Energy and Materials (CNPEM), Campinas, S\~ao Paulo 13083-970, Brazil}

\author{E. Granado}
\affiliation{``Gleb Wataghin'' Institute of Physics, University of Campinas - UNICAMP, Campinas, S\~ao Paulo 13083-859, Brazil}

\begin{abstract}
Two intriguing unresolved issues of iridate physics are the avoided metallization under applied pressure of undoped Sr$_2$IrO$_4$ and related materials, and the apparent absence of superconductivity under electron doping despite the similarity of the fermiology of these materials with respect to cuprates. Here, we investigate the crystal structure and lattice vibrations of Sr$_2$IrO$_4$ by a combined phonon Raman scattering and x-ray powder diffraction experiment under pressures up to 66 GPa and room temperature. Density functional theory (DFT) and {\it ab}-initio lattice dynamics calculations were also carried out. A first-order structural phase transition associated with an 8 \%\ collapse of the $c$-axis is observed at high pressures, with phase coexistence being observed between $\sim 40$ and 55 GPa. At lower pressures and still within the high-symmetry tetragonal phase, a number of lattice and phonon anomalies were observed, reflecting crossovers between isostructural competing states. A critical pressure of $P_1=17$ GPa is associated with the following anomalies: (i) a reduction of lattice volume compressibility and a change of behavior of the tetragonal $c/a$ ratio take place above $P_1$; (ii) a four-fold symmetry-breaking lattice strain associated with lattice disorder is observed above $P_1$; (iii) two strong Raman active modes at ambient conditions (at $\sim 180$ and $\sim 260$ cm$^{-1}$) are washed out at $P_1$; and (iv) an asymmetric Fano lineshape is observed for the $\sim 390$ cm$^{-1}$ mode above $P_1$, revealing a coupling of this phonon with electronic excitations. DFT indicates that the Ir$^{4+}$ in-plane canted magnetic moment is unstable against a volume compression, indicating that the phase above $P_1$ is most likely non-magnetic. Exploring the similarities between iridate and cuprate physics, we argue that these observations are consistent with the emergence of a rotational symmetry-breaking electronic instability at $P_1$, providing hints for the avoided metallization under pressure and supporting the hypothesis of possible competing orders that are detrimental to superconductivity in this family. Alternative scenarios for the transition at $P_1$ are also suggested and critically discussed. Additional phonon and lattice anomalies in the tetragonal phase are observed at $P_2=30$ and $P_3=40$ GPa, indicating further competing phases that are stabilized at high pressures. 

\end{abstract}

\maketitle

\section{Introduction}

\begin{figure*}
\includegraphics[width=0.8\textwidth]{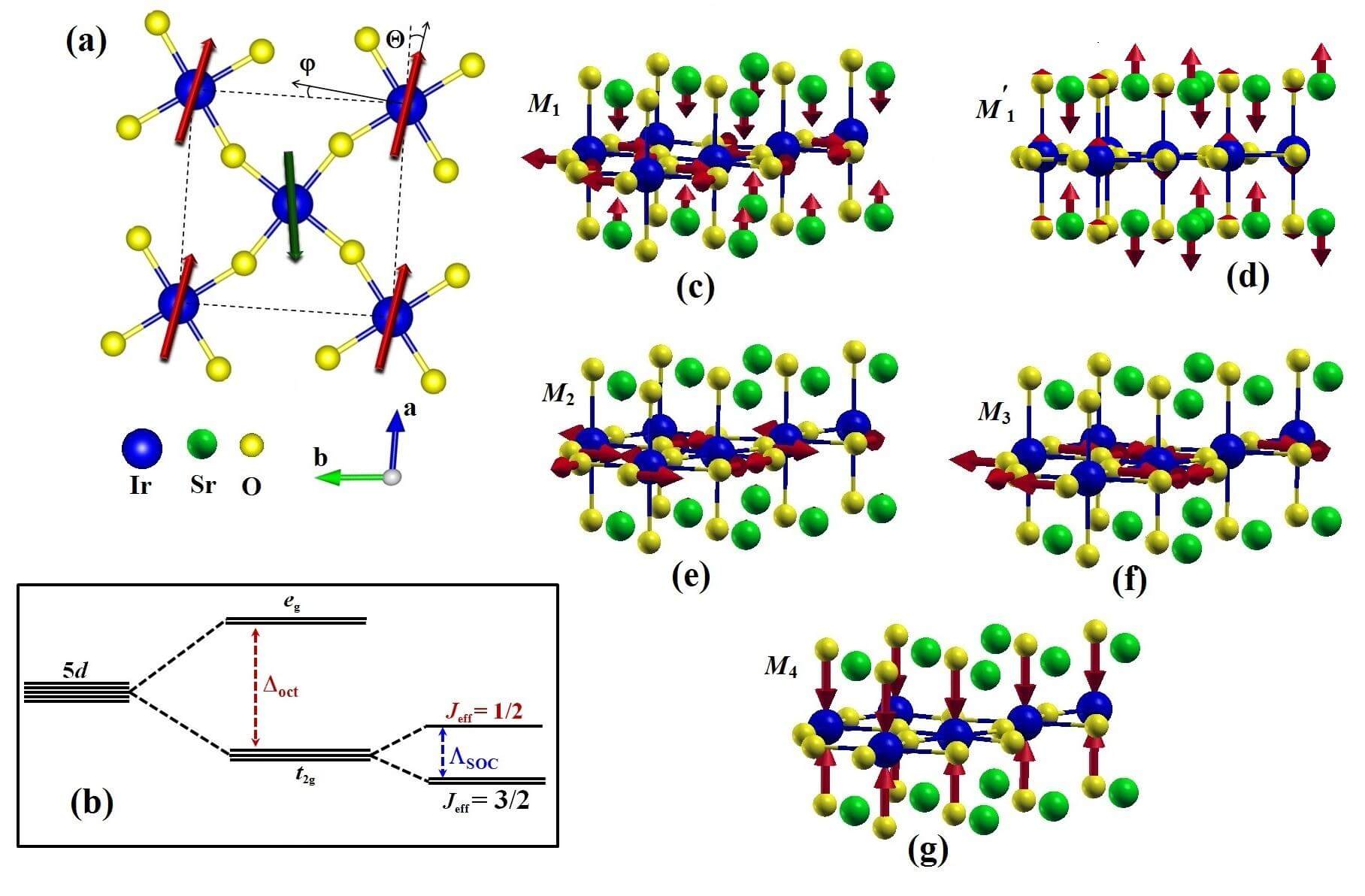}
\begin{quotation}
\caption{(a) $ab$-plane projection of the crystal and magnetic structures of Sr$_2$IrO$_4$ showing the octahedral rotation angle $\phi$ along the $c$-axis and the magnetic canting angle $\Theta$. (b) Atomistic energy level diagram of the Ir $5d$ shell. (c)-(g) Mechanical representations of modes $M_1/M_1'$ ($\sim 185$ cm$^{-1}$), $M_2$ ($\sim 270$ cm$^{-1}$), $M_3$ ($\sim 395$ cm$^{-1}$), and $M_4$ ($\sim 570$ cm$^{-1}$).}
\label{modes}
\end{quotation}
\end{figure*}

Interest on iridates and other $5d$ transition-metal oxides is steadily growing due to the renewed perception that new physics arises from the unique combination of strong spin-orbit coupling ($\Lambda_{SOC} \sim$ 0.5 eV), intermediate on-site Coulomb interaction ($U_{eff} \sim$ 1.5-2.0 eV), and large $5d$ spatial extension \onlinecite{Pesin,Rau,Caorev}. Some $5d$-based materials are on the verge of magnetism, while others are magnetic but show much reduced magnetic moments with respect to atomistic values. This area of research also gained significant momentum with the realization of a spin-orbit-entangled ground state with $J_{eff}=1/2$ in the Ruddlesden-Popper iridates Sr$_2$IrO$_4$ and Sr$_3$Ir$_2$O$_7$ \onlinecite{Kim,Kim2,Jackeli,Caorev}. This exotic state arises from the splitting of the $5d$ $t_{2g}$ levels into a lower four-fold ($J_{eff}=3/2$) and an upper two-fold ($J_{eff}=1/2$) level by the spin-orbit interaction [see Fig. \ref{modes}(b)]. For Ir$^{4+}$ ions, these levels are occupied by five electrons, leaving one electron in the degenerate $J_{eff}=1/2$ level that is further split by Mott and Stoner physics \onlinecite{Kim,Arita,Li}, leading to a magnetic insulating ground state with pseudospin-1/2 moments and relatively sharp valence and conduction bands. The magnetic interactions between Ir pseudospins were found to be predominantly Heisenberg-type \onlinecite{Jackeli,KimRIXS,Fujiyama,Kim3,Katukuri}.

The similarities between the crystal and electronic structures of Sr$_2$IrO$_4$ and the parent cuprate superconductor La$_2$CuO$_4$ led to the expectation that the former material could have the basic elements to show superconductivity under electron doping \onlinecite{Wang,Watanabe,Meng}. Indeed, it was found experimentally that electron-doped Sr$_2$IrO$_4$ shows a fermiology that reproduces in many aspects the caracteristics of cuprates \onlinecite{Kim_ARPES1,Yan,Kim_ARPES2}, however the observation of an actual superconducting state in this and related iridates is still missing. This suggests that electronically ordered phases in iridates may compete with superconductivity, in similarity to the reported charge stripes, charge density wave and spin density wave instabilities in cuprates \onlinecite{Valla,Wu,Chang,Ghiringhelli,Kawasaki,Birgeneau,Wakimoto,Wakimoto2} and nematic magnetic order in Fe-based parent superconductors \onlinecite{Fernandes}. In this direction, a recent resonant x-ray diffraction study indicated a spin-density wave state in (Sr$_{1-x}$La$_x$)$_2$IrO$_4$ [Ref. \onlinecite{SDW}]. Another variable that may play a key role is the substitutional disorder inherent to chemical doping, which may destabilize both superconductivity and competing electronically ordered states. In fact, the most conspicuous results obtained by angular-resolved photoemission spectroscopy and scanning tunelling spectroscopy linking iridates to cuprates were obtained in clean surface-doped samples by potassium deposition method \onlinecite{Kim_ARPES1,Yan,Kim_ARPES2}, rather than by bulk chemical substitution.

Application of an external pressure is naively expected to induce metallization in undoped iridates without introducing the disorder inherent to chemical doping, being a possible pathway to access cooperative electronic phenomena in this system. However, a metallic state is not achived up to at least 40 GPa in Sr$_2$IrO$_4$. Rather, the pressure-dependence of resistivity at low temperatures displays an $U$-shaped curve \onlinecite{Haskel,Zocco,Caorev}. For instance, at 50 K, the resistivity shows a drop of three orders of magnitude from ambient pressure up to $\sim 17$ GPa, remains nearly constant up to $\sim 30$ GPa, and then increases again above this pressure \onlinecite{Haskel,Caorev}. The reason for this avoided metallization, which is extensible to other related materials \onlinecite{Caorev}, remains as a major unresolved issue of iridate physics.

Structurally, parent Sr$_2$IrO$_4$ crystallizes in a tetragonal phase with space group $I$4$_1$/${acd}$, showing a rotation of the IrO$_6$ octahedra by $\phi = 11^\circ$ along the $c$-axis [Refs. \onlinecite{Huang,Crawford}. A canted magnetic structure with a weak ferromagnetic moment of 0.06-0.14 $\mu_B$/Ir is observed below $T_N=240$ K, where the canting angle $\Theta$ of pseudospins is locked to $\phi$ [see Fig. \ref{modes}(a)] \onlinecite{Kim,Jackeli,Crawford,Chikara,Cao,Kim2,Ye,Boseggia,Liu}. An x-ray spectroscopy study by Haskel {\it et al.} \onlinecite{Haskel} showed a magnetic circular dichroism signal associated with the weak ferromagnetic moment that disappears for $P > 17$ GPa. It was also shown that the Ir $L_3/L_2$ branching ratio reduces considerably with pressure with an anomaly at $\sim 30-40$ GPa \onlinecite{Haskel}, revealing a significant sensitivity of the Ir $5d$ electronic configuration to external pressures. 


A better understanding of the avoided metallization of Sr$_2$IrO$_4$ and related iridates under pressure may reveal competing ground states and provide insight into the lack of superconductivity in this family. In this work, the crystal structure and vibrational properties of this material are investigated by a combined x-ray powder diffraction (XRD) and  phonon Raman scattering experiment at room temperature and pressures up to 66.0 GPa, supported by density functional theory calculations. Supplementary lattice dynamics calculations are also performed and given in Appendix A. A first-order structural transition between tetragonal and monoclinic phases characterized by an 8 \%\ collapse of the $c$-axis length and a comparable expansion of the $b$ axis is observed at high pressures, with a phase coexistence region between 40 and 55 GPa. In addition, within the tetragonal phase, structural and phonon anomalies were observed at the characteristic pressures of 17, 30, and 40 GPa. The state between 17 and 30 GPa shows lattice strain associated with an incipient tetragonal-symmetry breaking instability. It is suggested that the observed lattice and phonon anomalies below 40 GPa mark transitions between competing electronic ground states, which may be the key to understand its avoided metallization under pressure and may shed light onto the missing superconductivity in the phase diagram of electron-doped iridates. Alternative scenarios that are also compatible with our results are critically discussed.

\section{Experimental and Computational Details}

\subsection{Experimental details}
	
The Sr$_2$IrO$_4$ powder sample employed in this work was prepared by a standard high temperature solid state reaction mechanism. High purity IrO$_2$ and SrCO$_3$ were mixed in stoichiometric ratio, calcined at 1100 $^{\circ}$C for 24 h and cooled to room temperature at the rate of 4 $^{\circ}$C/min. Laboratory x-ray diffraction measurements at ambient conditions were performed with a Bruker D2 Phaser diffractometer using Cu $K\alpha$ radiation. Magnetization measurements were taken with a Quantum Design Superconducting Quantum Interference Device (SQUID) magnetometer, showing a magnetic ordering transition temperature of 237 K (see Appendix B).

Pressure-dependent synchrotron x-ray diffraction (XRD) and Raman spectroscopy experiments were sequentially performed at each pressure at the X-ray Diffraction and Spectroscopy (XDS) beamline of the Brazilian Synchrotron Light Source (LNLS) \onlinecite{XDS}. Pressure was applied by diamond anvil cells (DAC) using Boehler-Almax-type ultra-low fluorescence diamonds with cullet diameter of 350 microns. The pressure transmitting medium was neon \onlinecite{neon} and gaskets were made of rhenium. The value of $P$ was obtained using the well-known ruby $R_1$ fluorescence line shift method \onlinecite{Ruby}.

The unpolarized Raman spectroscopy measurements under pressure were taken using a 532 nm diode laser, a 90 cm$^{-1}$ transition long-pass edge filter, a single stage 300 mm focal length Czerny-Turner spectrograph with 1800 gr/mm grating and a liquid nitrogen cooled CCD. The beam was focused and scattered light collected by a confocal setup. The laser beam focal spot was $\sim 40$ $\mu$m and the laser power was kept below 20 mW. The instrumental linewidth given by this setup is 3 cm$^{-1}$.

XRD intensities were collected using a monochromatic beam with $\lambda=0.62023$ \AA\, calibrated using a LaB$_6$ standard, and a focal spot size of $90 \times 40$ $\mu$m obtained with a Rh cylindrical collimating mirror, a LN2-cooled double flat Si(111) crystal monochromator, a Rh toroidal focusing mirror and a Kirkpatrick-Baez mirror. The diffracted Debye rings were detected in transmission geometry by a 2D-detector (Rayonix MX225 with 73 mm pixel size) placed 25 cm away from the sample; the ring intensities were integrated to yield a conventional $I$ versus $2 \theta$  plot. Useful diffraction data were obtained up to $2 \theta = 28^{\circ}$, corresponding to Bragg reflections with interplanar distance $d > 1.28$ \AA. The maximum pressure was 66.0 GPa.

Preliminary XRD + Raman scattering data were collected in a separate beamtime period, using a setup similar to the above but with the following differences: (i) the calibrated wavelength was $\lambda=0.62304$ \AA; (ii) the Kirkpatrick-Baez mirror was not employed to focus the x-ray beam, and a circular beam of diameter $\sim 100$ $\mu$m was defined by an x-ray pinhole placed close to the pressure cell; (iii) the pressure cell had a smaller angular aperture ($2 \theta = 21^{\circ}$, corresponding to $d > 1.7$ \AA); (iv) the maximum pressure was 45.0 GPa; and (v) a significant luminescence signal from diamond was observed as a baseline in the Raman spectra. These preliminary data are presented in Appendices C and D, being consistent with the data shown in the main text.

\subsection{Computational details}

Density Functional Theory (DFT) was employed as implemented in the QUANTUM ESPRESSO suite \onlinecite{QE}. Plane-wave self-consistent field calculations were extensively performed using the PWscf core package, while lattice dynamics calculations were carried out using the specialized PHonon package in addition to PWscf.

Non-magnetic atomic-position relaxation and lattice dynamics calculations were carried out with the Scalar Relativistic Density-Functional Perturbation Theory under the Generalized Gradient Approximation (GGA), using the Perdew, Burke, and Enzerhof (PBE) exchange-correlation potential \onlinecite{Perdew} and ultrasoft pseudopotentials \onlinecite{pseudopotentials}. The energy cuttoffs for the wavefunctions and charge density were 70 and 560 Ry, respectively, and a $4 \times 4 \times 4$ Monkhorst-Pack grid of $k$-points was employed. The experimental lattice parameters obtained in this work at several pressures were used as input for the atomic-position relaxation calculations, while the reported experimental crystal structure at $T=13$ K \onlinecite{Crawford} was used as input for the lattice dynamics calculations.

Additional electronic structure calculations with a canted magnetic structure were performed under the GGA and projector augmented wave formulation with a PBE exchange-correlation potential \onlinecite{pseudopotentials2}. Here, spin-orbit interaction was taken into account and Ir $5d$ on-site Coulomb interaction $U=2.3$ eV and effective exchange parameter $J=0.3$ eV were included \onlinecite{Himmetoglu}. The non-colinear magnetic moments were obtained by separating the Ir ions into two species, corresponding to the distinct IrO$_6$ octahedral orientations [see Fig. \ref{modes}(a)]. The initial Ir magnetization was chosen to be in-plane [see Fig. \ref{modes}(a)]. Here, the energy cuttoffs for the wavefunctions and charge density were 56 and 319 Ry, respectively, and a $6 \times 6 \times 2$ Monkhorst-Pack grid of $k$-points was employed. For these calculations, we used as input the lattice parameters at 2.5 and 9.6 GPa experimentally obtained in this work and the relaxed atomic positions obtained from the atomic-position relaxation calculations detailed above.

Rietveld refinements and Le Bail fits using XRD data were performed with the GSAS+EXPGUI suite \onlinecite{GSAS}. The mechanical representations of the lattice vibrations were drawn using the program XCrySDen \onlinecite{xcrysden}, while Fig. \ref{modes}(a) was prepared with the aid of the software VESTA \onlinecite{VESTA}.

\section{Results and Analysis}

\subsection{X-ray diffraction}

\begin{figure}
	\includegraphics[width=0.45\textwidth]{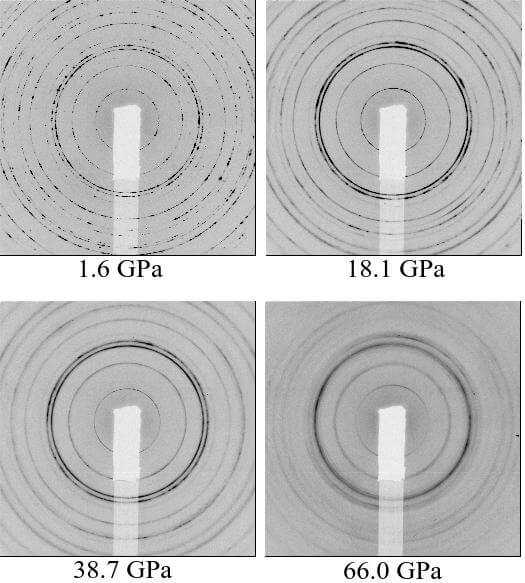}
	\begin{quotation}
		\caption{Raw images of x-ray powder diffraction data at selected pressures.}
		\label{XRD_Images}
	\end{quotation}
\end{figure}

\begin{figure}
	\includegraphics[width=0.45\textwidth]{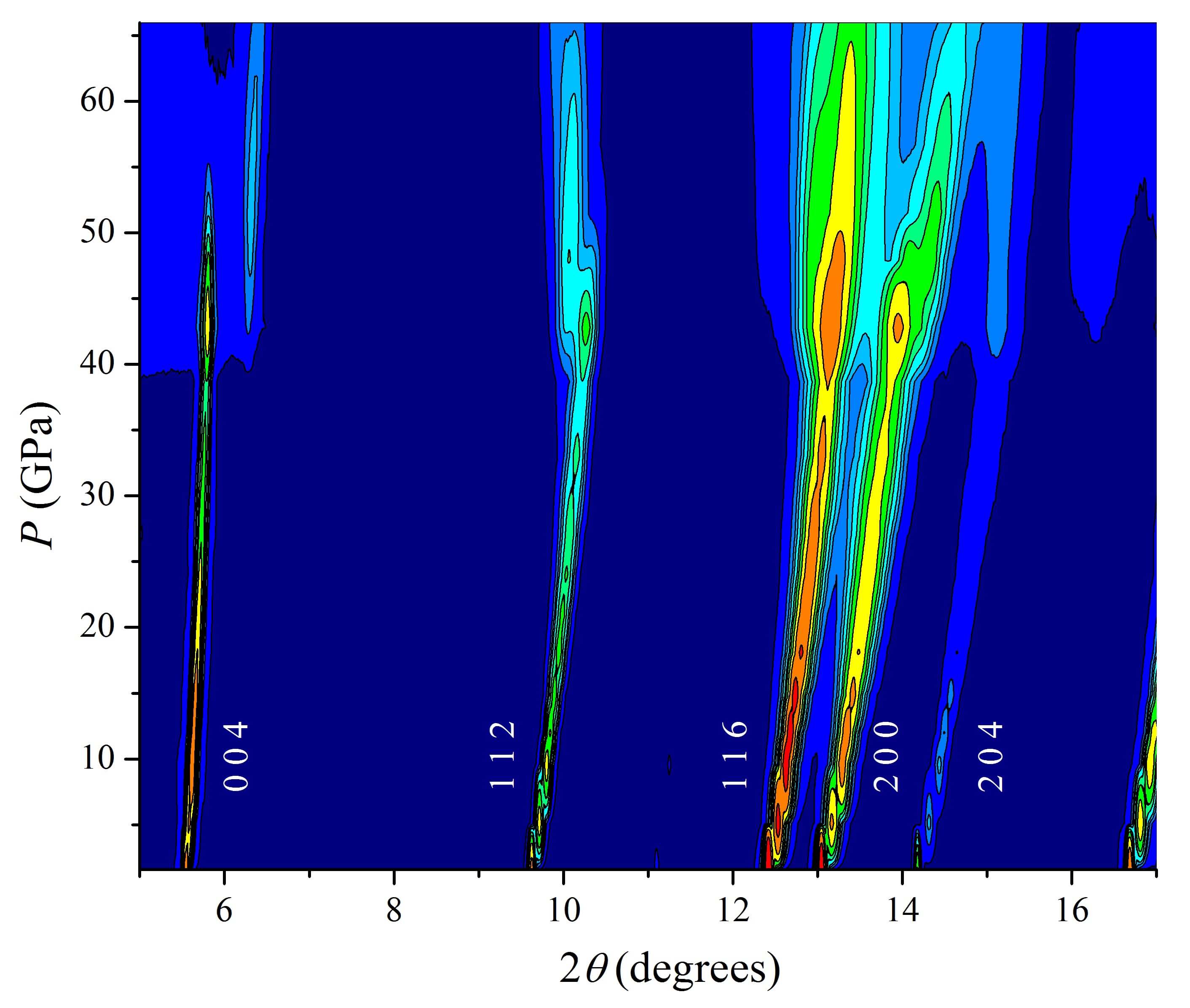}
	\begin{quotation}
		\caption{Contour plot illustrating the evolution of the x-ray diffraction profiles with applied pressure in a selected angular region ($\lambda=0.62023$ \AA). The reflection {\it hkl} indexes are indicated.}
		\label{XRD_contour}
	\end{quotation}
\end{figure}

\begin{figure}
	\includegraphics[width=0.35\textwidth]{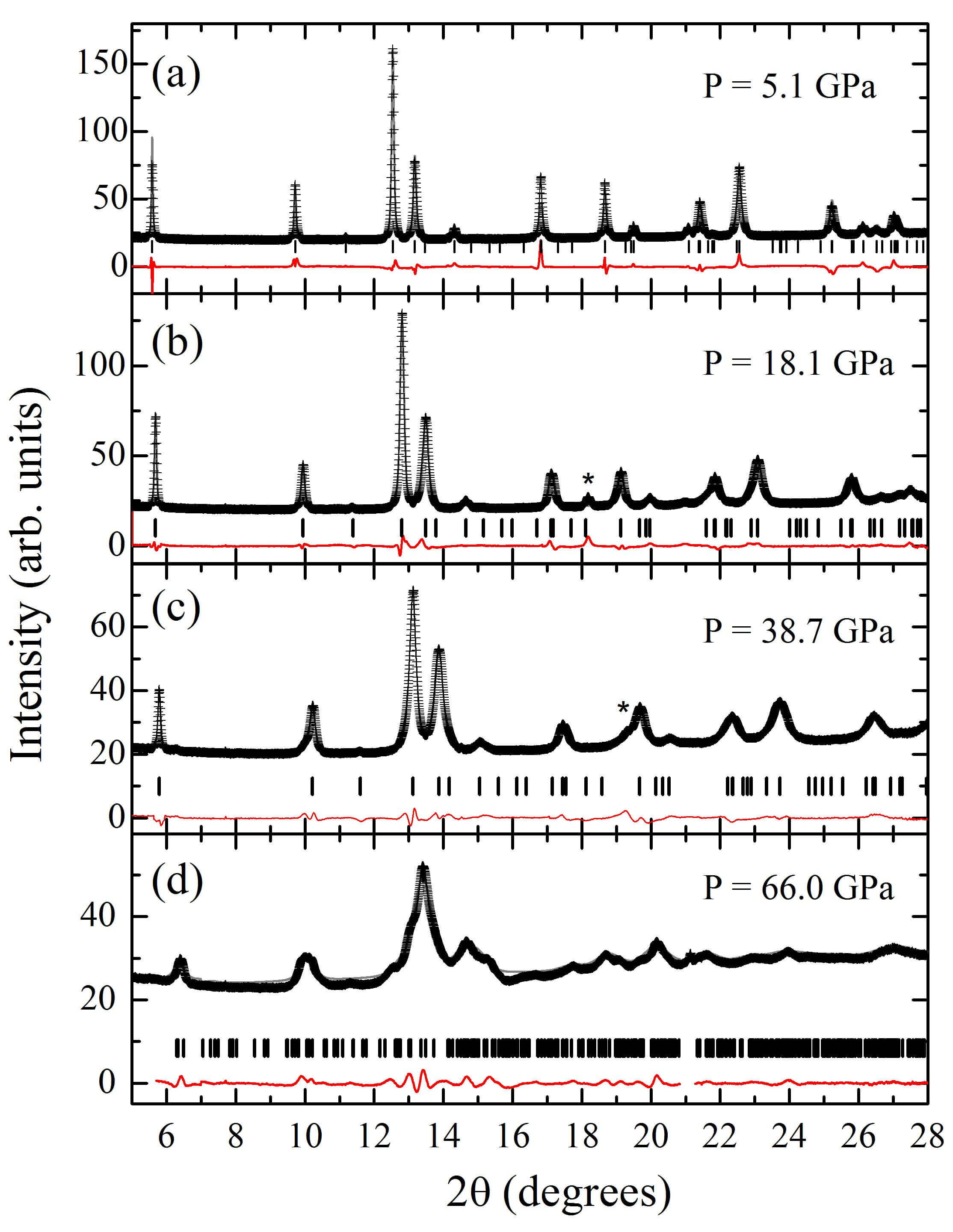}
	\includegraphics[width=0.40\textwidth]{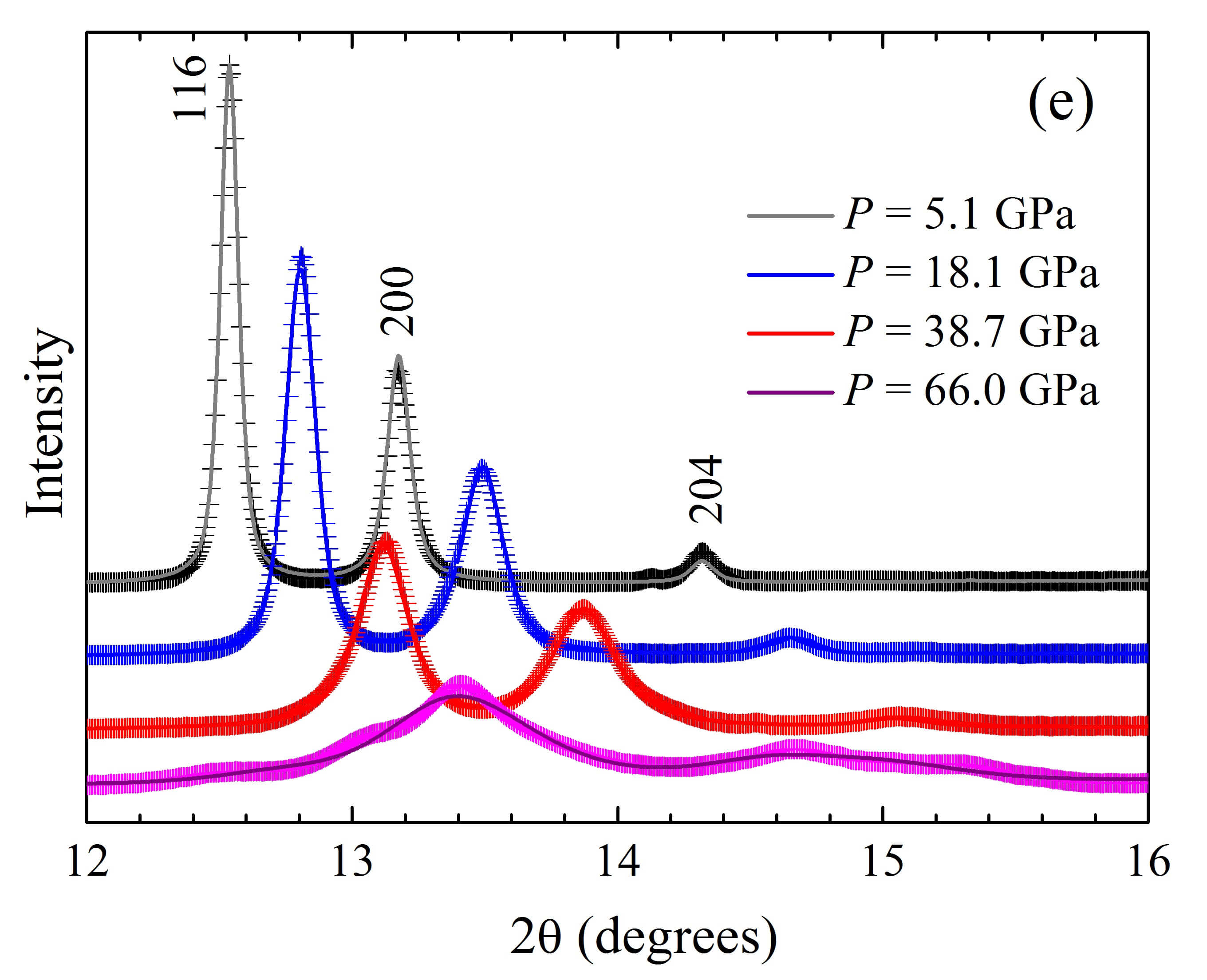}

	\begin{quotation}
		\caption{(a-d) Observed (crosses) and calculated (solid lines) x-ray diffraction intensities at $P=5.1$ (a), 18.1 (b), 38.7 (c), and 66.0 GPa (d), taken with $\lambda=0.62023$ \AA. The difference curves are given as solid red lines and the short vertical lines mark the expected reflection positions. The calculated intensities and peak positions  refer to a Rietveld refinement under the tetragonal space group $I4_1/acd$ in panels (a-c) and to a Le Bail fit under the monoclinic space group $P_2$ in panel (d) (see text). The star symbols in (b) and (c) mark a Bragg peak from Neon. (e) Zoom out of (a-d) covering the (tetragonal) 116, 200 and 204 reflections.}
		\label{XRD_Rietveld}
	\end{quotation}
\end{figure}

\begin{figure}
	\includegraphics[width=0.45\textwidth]{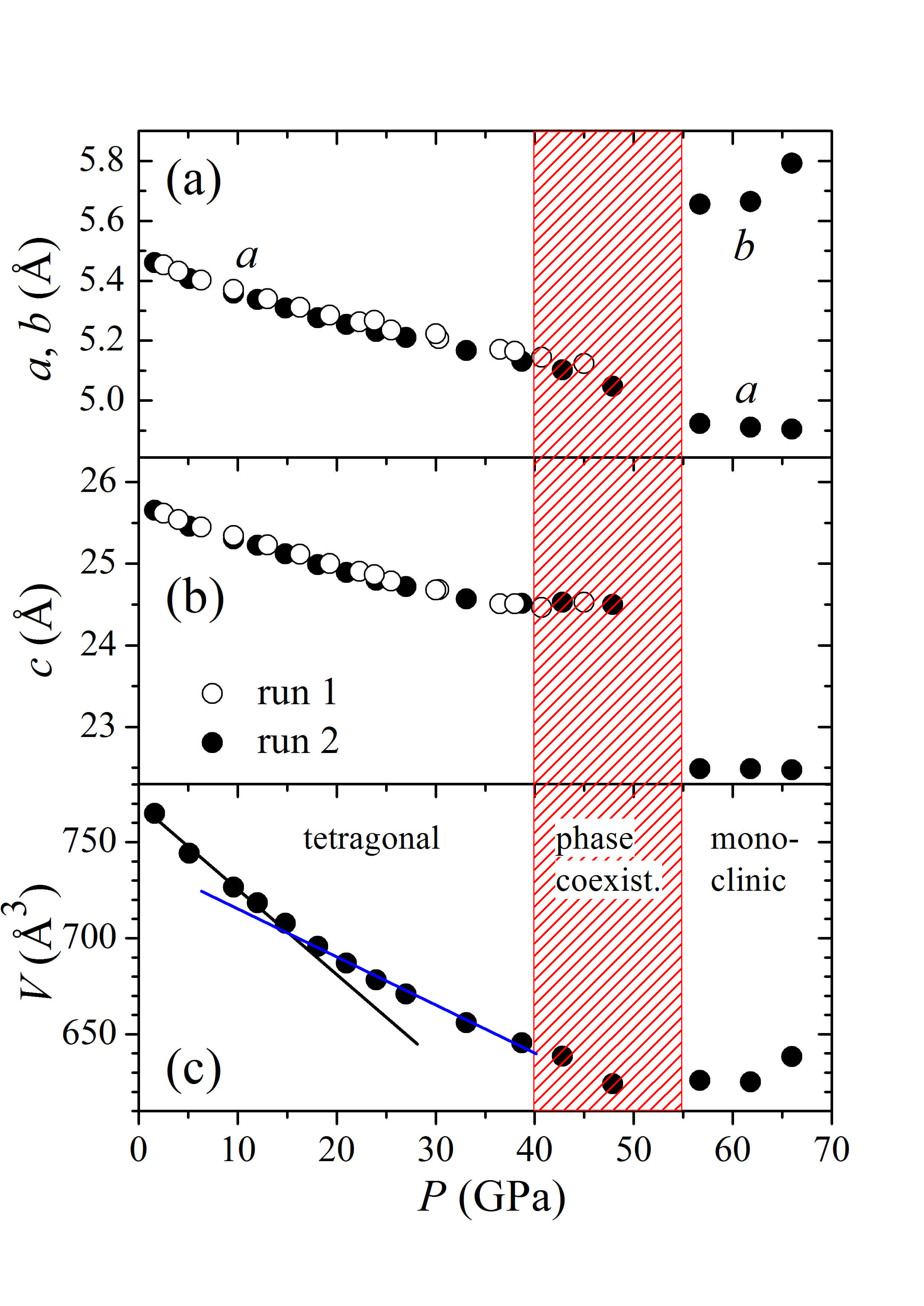}
	\begin{quotation}
		\caption{Pressure-dependence of tetragonal $a$ and monoclinic $a$ and $b$ lattice parameters (a), $c$ lattice parameter (b) and unit-cell volume (c) obtained from the fits illustrated in Figs. \ref{XRD_Rietveld}(a-d). Open symbols represent data taken in a preliminary run (see also Appendix C). The dashed area marks the pressure region showing phase coexistence (see also Fig. \ref{XRD_contour}). Solid straight lines in (c) are guides to the eyes.}
		\label{XRD_latticepar}
	\end{quotation}
\end{figure}

\begin{figure}
	\includegraphics[width=0.45\textwidth]{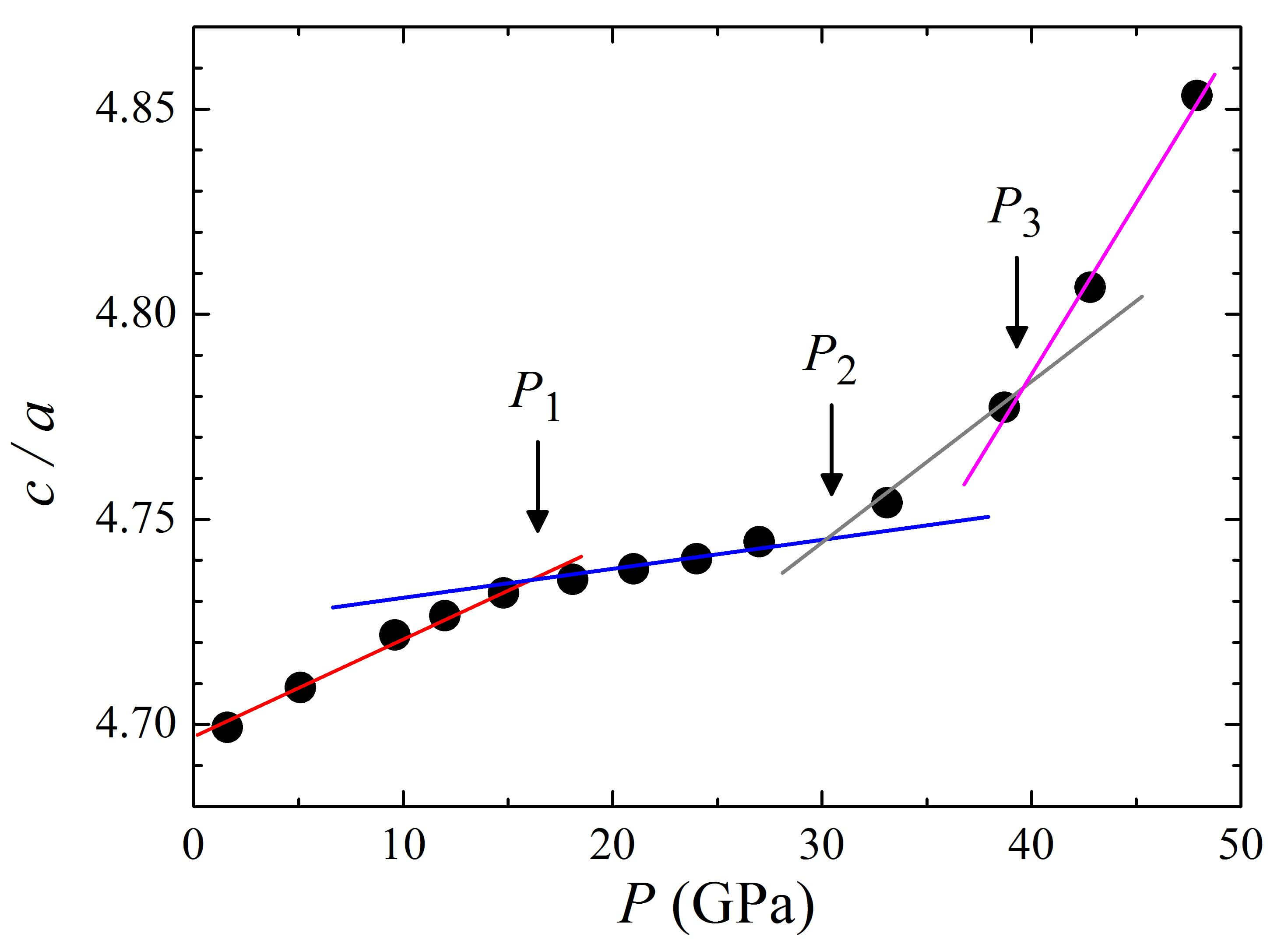}
	\begin{quotation}
		\caption{Pressure-dependence of $c/a$ ratio. Solid lines are guides to the eyes. Vertical arrows mark the reference pressures $P_1=17$, $P_2=30$, and $P_3=40$ GPa.}
		\label{XRD_c_over_a}
	\end{quotation}
\end{figure}

\begin{figure}
	\includegraphics[width=0.45\textwidth]{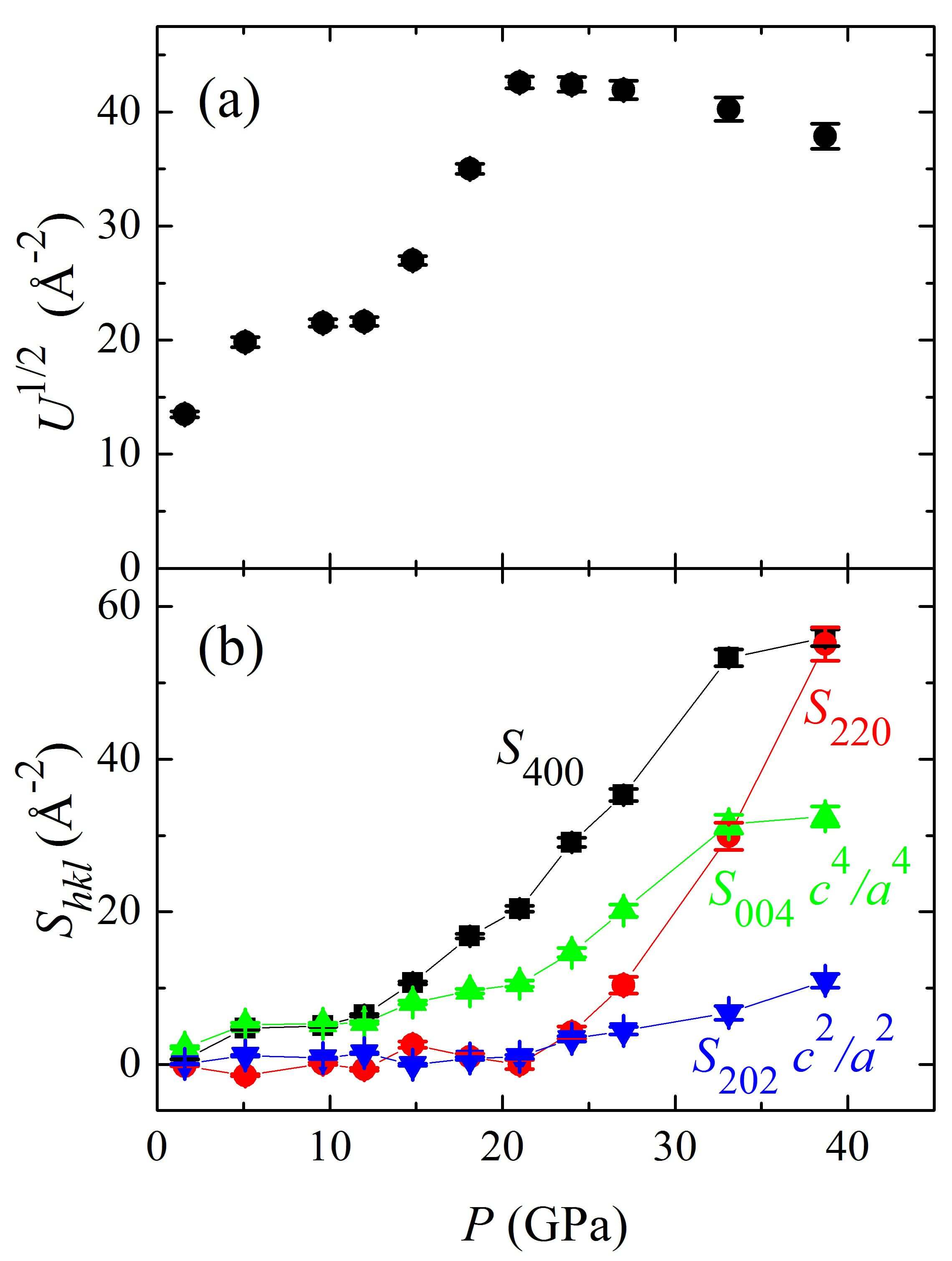}
	\begin{quotation}
		\caption{Pressure-dependence of isotropic Gaussian (a) and anisotropic Lorentzian (b) strain broadening parameters of Bragg peaks (run 2).}
		\label{XRD_strain}
	\end{quotation}
\end{figure}

Figure \ref{XRD_Images} shows raw $2D$-detector XRD images at selected pressures. At low pressures, significant graininesses of the Debye rings are noticed, due to the rather limited amount of material inside the pressure cell. The grain statistics is improved with increasing pressures, which is ascribed to Bragg peak broadening (see below) that increases the number of crystallites in Bragg condition for each reflection. Figure \ref{XRD_contour} shows a contour plot illustrating the pressure-dependence of the $I(2 \theta$) diffractograms of Sr$_2$IrO$_4$ in a selected angular interval, obtained by averaging out the ring intensities of the images shown in Fig. \ref{XRD_Images}. With increasing pressures up to $\sim 40$ GPa, the Bragg peaks move to higher angles, consistent with lattice compression. A first-order structural phase transition clearly takes place at higher pressures, with phase coexistence being observed between $\sim 40$ and $55$ GPa. The high-pressure phase shows a significant displacement of the $004$ reflection to higher diffraction angles, revealing a collapse of the $c$-axis. The $112$ reflection moves to lower angles, indicating an {\it ab}-plane expansion. 

Figures \ref{XRD_Rietveld}(a)-\ref{XRD_Rietveld}(d) show the full x-ray diffraction profiles at selected pressures. For $P<40$ GPa, Rietveld refinements were performed under the tetragonal $I4_1/acd$ space group using the atomic parameters at ambient conditions reported in Ref. \onlinecite{Crawford} [see Figs. \ref{XRD_Rietveld}(a)-\ref{XRD_Rietveld}(c)]. On the other hand, the highly overlapping broad peaks at high pressures did not allow us to reach a satisfactory structural model for the high-pressure phase. A Le Bail fit for a low-symmetry monoclinic unit cell (space group $P_2$) was then performed to extract the lattice parameters [see Fig. \ref{XRD_Rietveld}(d)]. The pressure-dependence of the $a$, $b$, and $c$ lattice parameters and unit-cell volume of the low- and high-pressure phases are given in Figs. \ref{XRD_latticepar}(a-c). As anticipated above, a collapse of 8 \% in the $c$-parameter is seen for the high pressure phase, which is partly compensated by a large increment of the $b$-parameter. Similar trends for the lattice parameters were seen at the tetragonal-monoclinic transition of Sr$_3$Ir$_2$O$_7$ at 54 GPa \onlinecite{Donnerer}, suggesting a common mechanism for the structural transitions of both compounds.

Besides the structural phase transition observed between 40 and 55 GPa, the detailed pressure-dependence of the tetragonal phase also shows an interesting behavior. For instance, it can be seen in Fig. \ref{XRD_latticepar}(c) that a significant change in the volume compressibility takes place at $P_1 \sim 17$ GPa. A similar effect was reported for Sr$_3$Ir$_2$O$_7$ at $\sim 14$ GPa \onlinecite{Zhao}. In addition, Fig. \ref{XRD_c_over_a} shows the $c/a$ ratio, revealing that the unit cell response to pressure is anisotropic. In fact, the tetragonal $c/a$ ratio increases from 4.70 at low pressures to 4.85 at $P=48.1$ GPa. The cell elongation increases steeply up to $P_1$, remains nearly constant between $P_1$ and $P_2 \sim 30$ GPa, and increases again above $P_2$. For pressures above $P_3 \sim 40$ GPa, the tetragonal cell elongates at a significantly higher rate. Such additional cell elongation at high pressures cannot be trivially understood as a precursor of the structural phase transition, since the high-pressure phase has a collapsed $c$-parameter, contrary to the tendency of the tetragonal phase with increasing pressures. This strikingly rich behavior of the tetragonal phase is suggestive of a non-trivial lattice relaxation with increasing pressures, possibly as a response of electronic phase transitions.

Significant broadening of the Bragg peaks is observed with increasing pressures. As illustrated in Fig. \ref{XRD_Rietveld}(e) by the distinct widths of the $116/200$ nearby Bragg peaks, such peak broadening is anisotropic. In our Rietveld refinements, the peak lineshapes and widths were well modeled by a pseudo-Voigt function, with the Gaussian width being given by $\sigma = (U)^{1/2} tan \theta$, where $U$ is an isotropic strain parameter, and the Lorentzian broadening coefficient $\gamma$ being given by $\Gamma = \gamma_S d^2 tan \theta$, where $\gamma_S^2 = S_{400} (h^4 + k^4) + S_{004} l^4 + 3 [S_{220} h^2k^2 + S_{202}(h^2l^2+k^2l^2)]$ is an anisotropic strain parameter \onlinecite{GSAS,Stephens}. The parameters $U$, $S_{400}$, $S_{004}$, $S_{220}$, and $S_{202}$ above were freely refined. Figures \ref{XRD_strain}(a) and \ref{XRD_strain}(b) show the Gaussian and Lorentzian broadening parameters, respectively. The isotropic Gaussian component increases significantly between $\sim 15$ and 20 GPa, remaining stable at higher pressures. As seen in Fig. \ref{XRD_strain}(b), the Lorentzian broadening is highly anisotropic and dominated by the $S_{400}$ term between $\sim 15$ and 40 GPa. The relatively sharp jump observed in the Gaussian broadening term betwen 15 and 20 GPa is not observed for the Lorentzian broadening parameters. A detailed analysis of the pressure dependence of $S_{400}$ indicates two characteristic pressures $P_1=17$ GPa and $P_2=30$ GPa where a change of behavior is noticed [see Fig. \ref{XRD_strain}(b)], which coincides with the reference pressures where a change of behavior was found in the volume compressibility and $c/a$ [see Figs. \ref{XRD_latticepar}(c) and \ref{XRD_c_over_a}]. Also, $S_{220}$ remains insignificant at low pressures and experiences a drastic increase above $\sim P_2$ [see Fig. \ref{XRD_strain}(b)].  Unfortunately, it was not possible to extract reliable broadening parameters for the tetragonal phase above 40 GPa due to the large Bragg peak overlap caused by tetragonal-monoclinic phase coexistence. 

\subsection{Raman scattering}

\begin{figure}
	\includegraphics[width=0.5\textwidth]{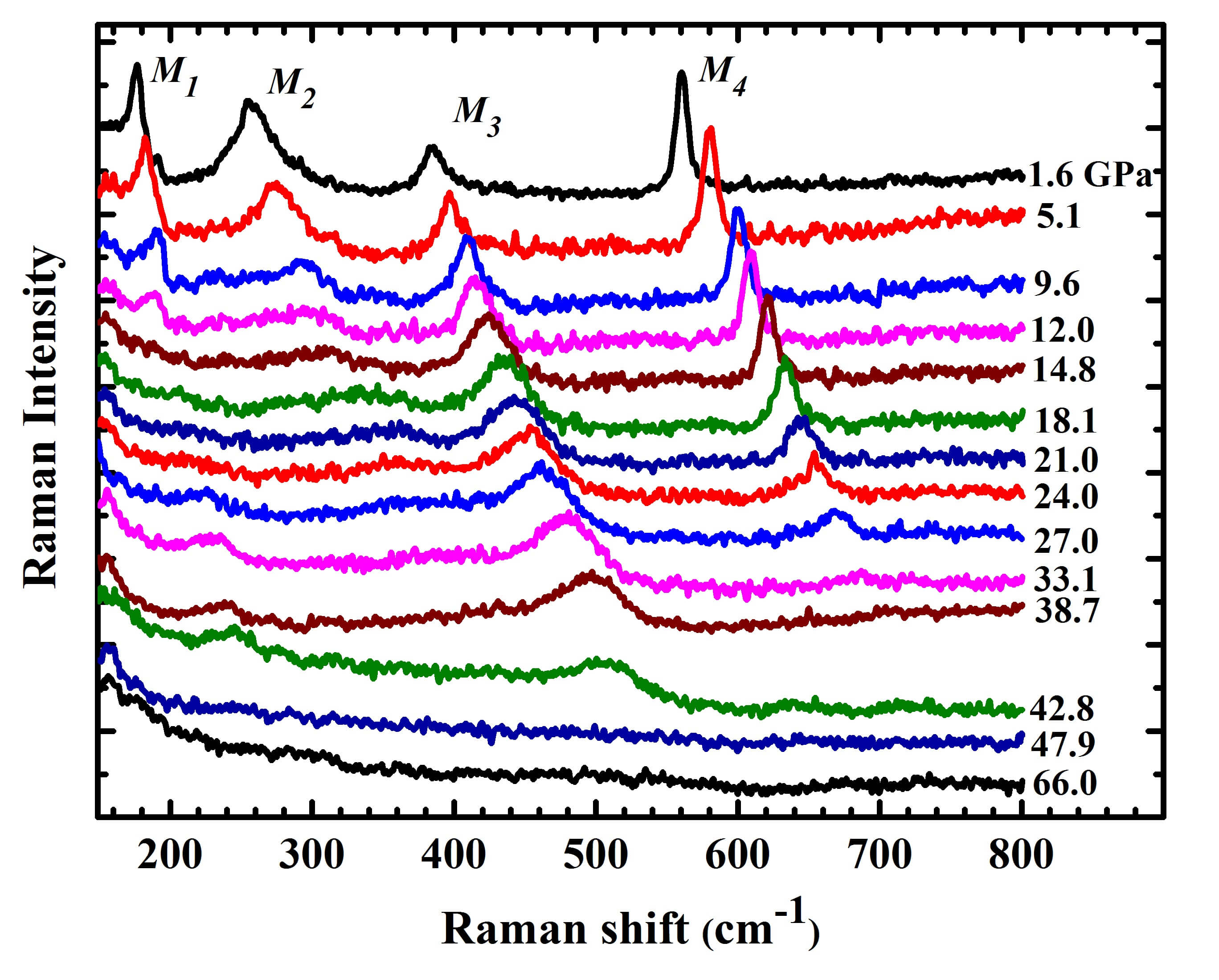}

	\begin{quotation}
		\caption{Raman spectra of Sr$_2$IrO$_4$ at room temperature at several pressures. The phonon peaks associated to modes $M_1$ and $M_2$-$M_4$ represented in Figs. \ref{modes}(c-g) are indicated.}
		\label{spectra}
	\end{quotation}
\end{figure}

\begin{figure}
	\includegraphics[width=0.5\textwidth]{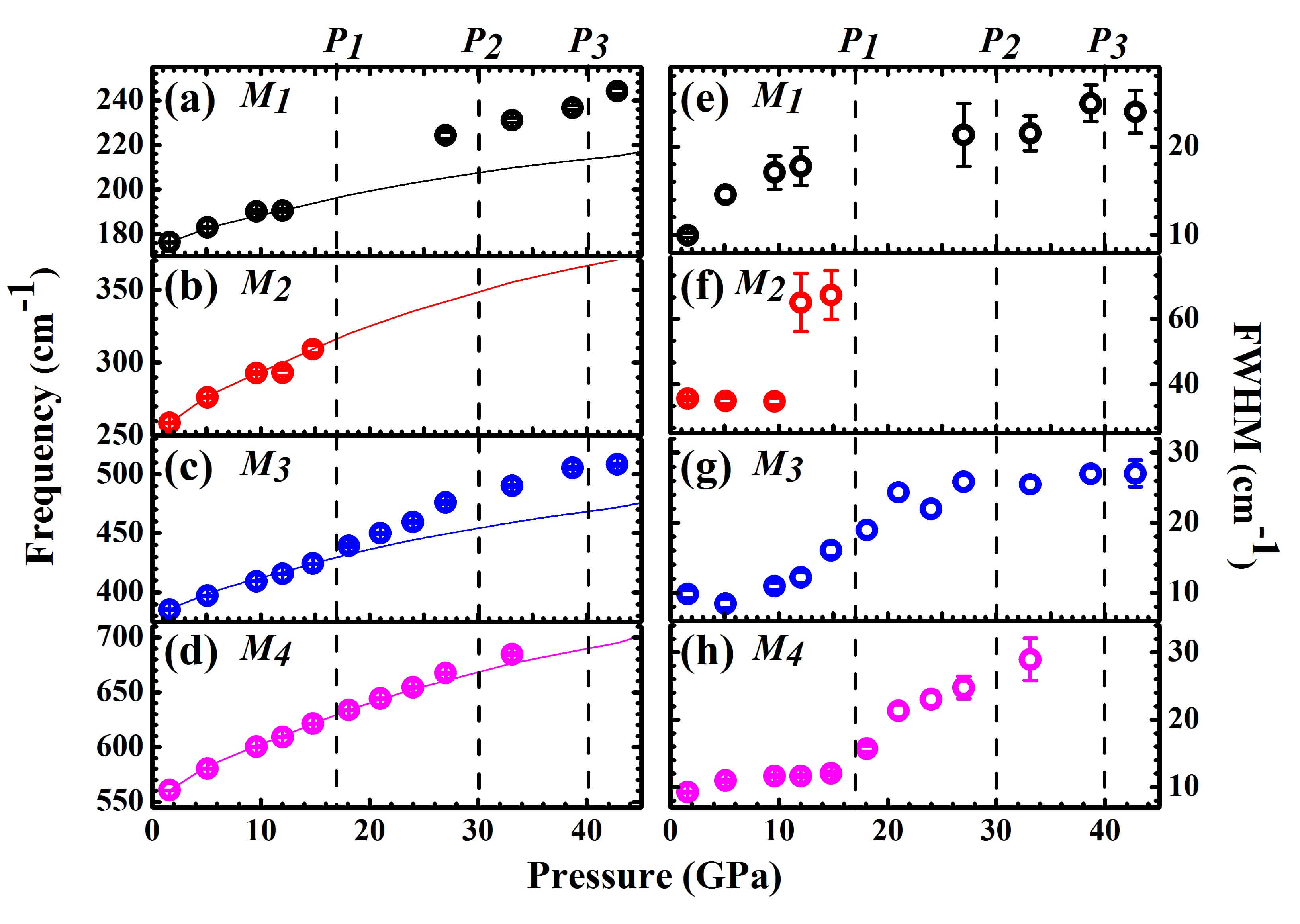}
	\includegraphics[width=0.5\textwidth]{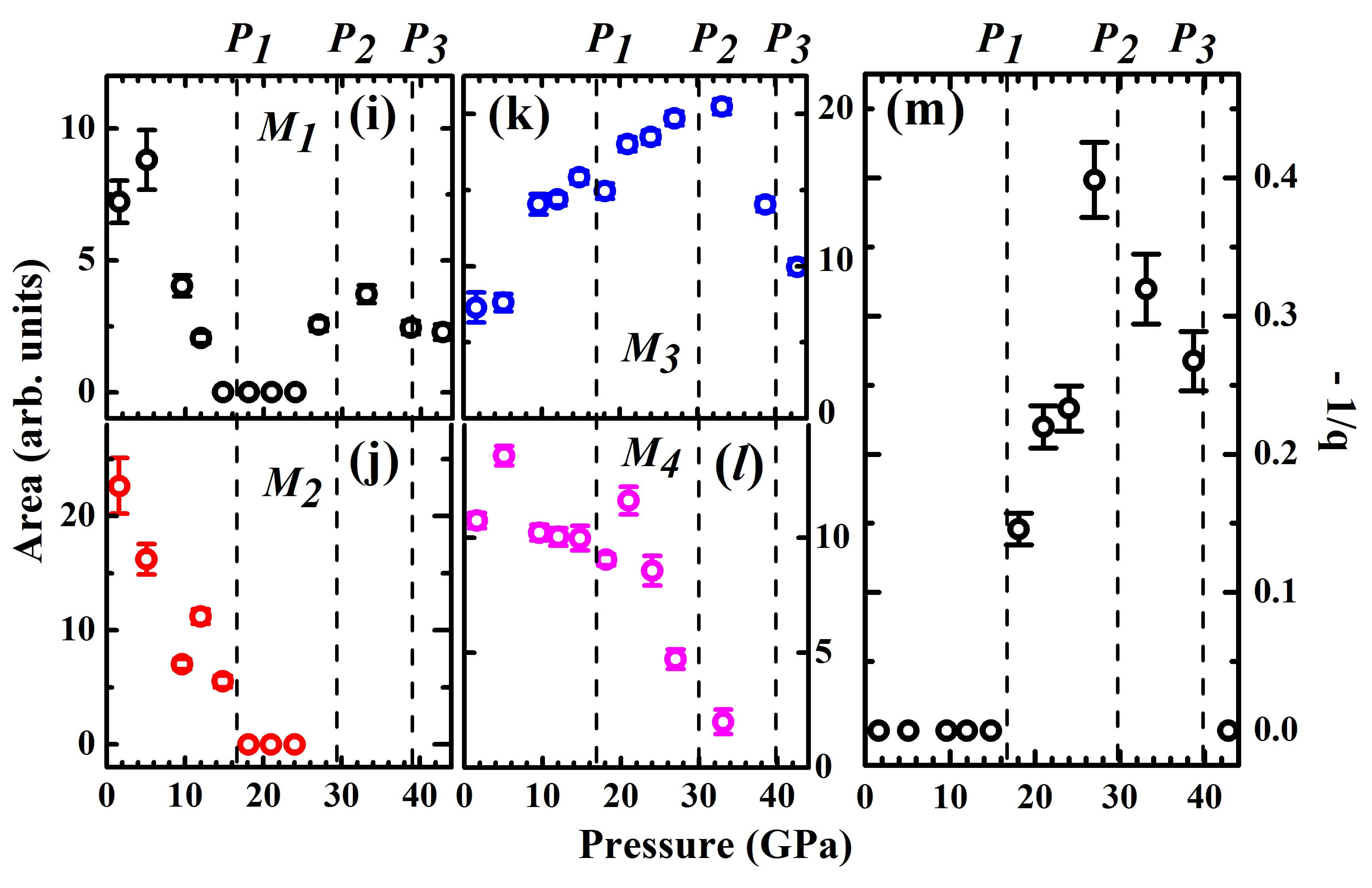}
	\begin{quotation}
		\caption{Pressure dependence of frequency (a-d), full width at half maximum (FWHM, e-h) and integrated area (i-l) of phonon Raman peaks $M_1$-$M_4$ (symbols). The vertical dashed lines mark the pressures $P_1 = 17$ GPa, $P_2=30$ GPa and $P_3=40$ GPa where phonon anomalies are noticed. The solid lines in (a-d) represent scalings to the unit cell volume according to the Gr\"uneisen's law (see text). (m) Electron-phonon coupling strength (-1/q) of mode $M_3$ as a function of pressure, obtained from its Fano lineshape asymmetry (see text).}
		\label{fitresults}
	\end{quotation}
\end{figure}
The Raman spectra of Sr$_2$IrO$_4$ at several pressures are shown in Fig. \ref{spectra}. A rich pressure-dependence of the Raman spectrum is observed, even below 40 GPa where no structural phase transition was detected by XRD (see above). At $P=1.6$ GPa and room temperature, four peaks at 180, 260, 390, and 560 cm$^{-1}$, labeled as $M_1-M_4$, dominate the spectrum. With increasing pressures, peaks $M_1$ and $M_2$ weaken and visually disappear above 14.8 GPa; $M_1$ seems to reemerge above 27.0 GPa, being then observed up to 42.8 GPa. $M_3$ is continuously observed up to $42.8$ GPa and $M_4$ disappears above 33.1 GPa. For $P \geq 47.9$ GPa, the phonon bands nearly disappear within our sensitivity, possibly marking the structural phase transition observed by XRD (see above). 

Figures \ref{fitresults}(a-l) show the $P$-dependence of $M_1-M_4$ frequencies (a-d), linewidths (e-h), and integrated areas (i-l). It is interesting to note that the anomalies in phonon intensities seem to occur at the same reference pressures $P_1$, $P_2$, and $P_3$ where anomalies in XRD data are observed (see above). In fact, modes $M_1$ and $M_2$ weaken with pressure and disappear at $\sim P_1$; $M_1$ reappears above $\sim P_2$; $M_3$ is enhanced up to $\sim P_2$ where it begins to weaken; finally, $M_4$ remains with a nearly constant intensity up to $\sim P_1$, weakens above this pressure and fades away above $\sim P_2$.

All observed modes harden and broaden with increasing $P$. The solid lines in Figs. \ref{fitresults}(a)-\ref{fitresults}(d) are scalings of the phonon frequencies to the Gr\"uneisen's law, $\Delta \omega / \omega \propto \Delta V / V$. Modes $M_2$ and $M_4$ follows this law in the pressures ranges where they are observed, while modes $M_1$ and $M_3$ show an anomalous hardening above $P_1$. Concerning the linewidths, modes $M_3$ and $M_4$ broaden above $P_1$, resembling very much the pressure-dependence of the $S_{400}$ Lorentzian strain broadening parameter [see also Fig. \ref{XRD_strain}(b)].

It is interesting to note that $M_3$ shows an asymmetric profile between $P_1$ and $P_3$, being fitted with a Fano lineshape $I(\omega)=I_0(q+\epsilon)^2/(1+\epsilon^2)$, where $I_0$ is the intensity, $q$ is the asymmetry parameter and $\epsilon \equiv (\omega-\omega_0)/\Gamma$, with $\omega_0$ and $\Gamma$ being the phonon frequency and linewidth, respectively (Ref. \onlinecite{Fano}, see also Appendix D). The pressure-dependence of the electron-phonon coupling strength $-1/q$ of this mode is given in Fig. \ref{fitresults}(m), being zero below $P_1$ and above $P_3$ within our resolution and maximum at $P_2$. The other modes could be well fit by symmetric Lorentzian lineshapes within our statistics.

\subsection{Density functional theory}

\begin{figure}
	\includegraphics[width=0.4\textwidth]{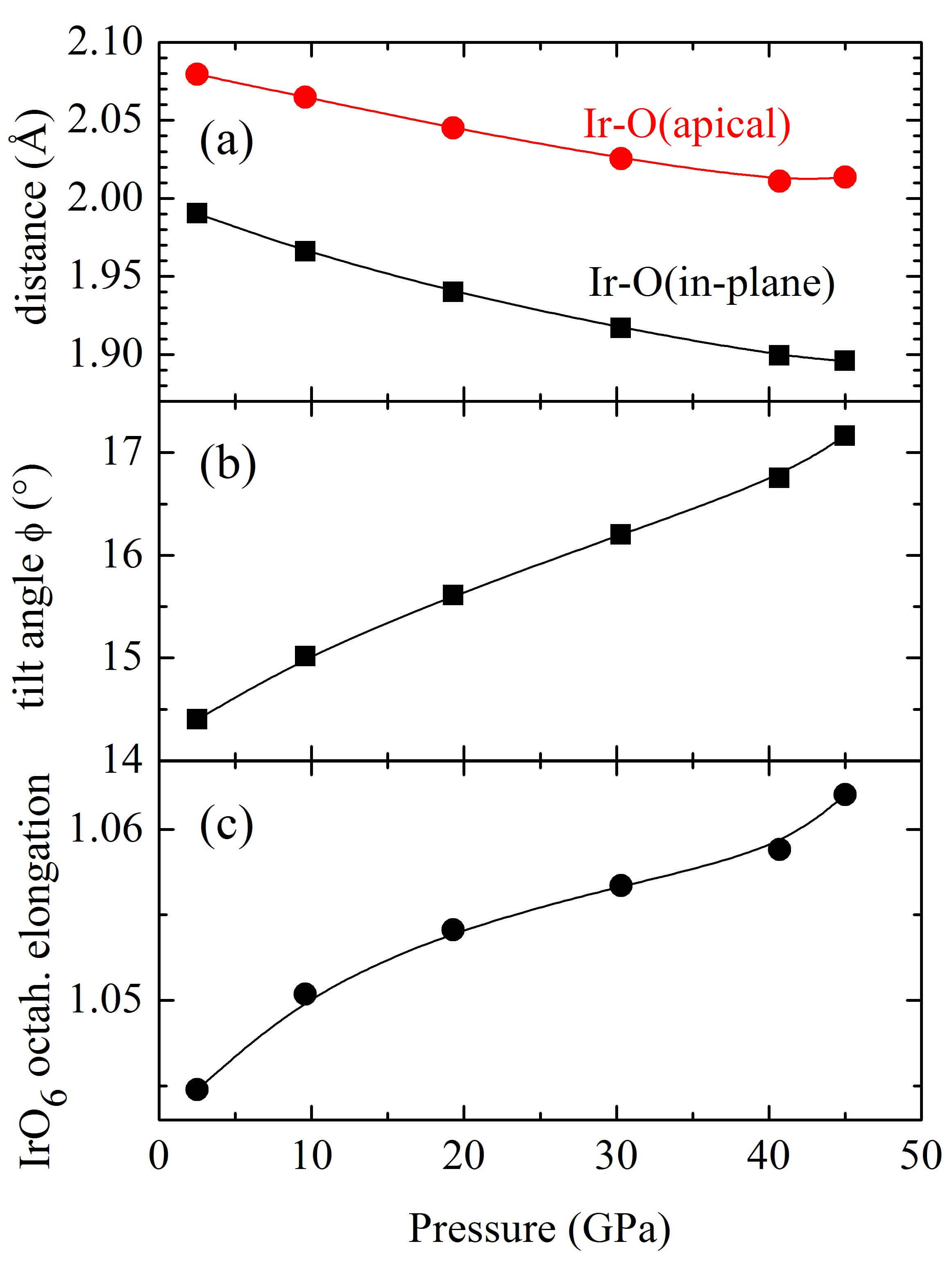}
	\begin{quotation}
		\caption{Calculated pressure dependence of Ir-O(in-plane) and Ir-O(apical) bond distances (a), Ir-O-Ir tilt angle $\phi$ (b), and tetragonal elongation of the IrO$_6$ octahedra Ir-O(apical)/Ir-O(in-plane) (c), obtained by density functional theory calculations using the experimental lattice parameters as input.}
		\label{DFT}
	\end{quotation}
\end{figure}

The pressure-dependencies of the Ir-O bond distances and in-plane bond angle $\phi$ [see Fig. \ref{modes}(a)] are desirable. Ideally, such information might be obtained experimentally from Rietveld fits of the diffraction data [Figs. \ref{XRD_Rietveld}(a-c)], however the relatively poor overall fittings due to grain statistics and preferred orientation, as well as the limited diffraction angle interval, did not allow for a reliable experimental determination of the oxygen atomic positions. Density functional theory calculations were then performed to obtain the relaxed bond distances and bond angles of the tetragonal phase as a function of pressure, using the experimental lattice parameters as input. Figure \ref{DFT}(a) shows the pressure-dependence of the calculated Ir-O(in-plane) and Ir-O(apical) bond lengths, while Fig. \ref{DFT}(b) shows the Ir-O-Ir tilt angle $\phi$. The calculated $\phi$ extrapolated to ambient pressure is $\sim 14^{\circ}$, in accordance with a previous calculation \onlinecite{Liu} but larger than the experimental value $\phi=11.5^{\circ}$ \onlinecite{Crawford}. It can be seen that the Ir-O bond lengths are reduced and $\phi$ is increased significantly under compression up to 45 GPa. Also, the tetragonal elongation of the IrO$_6$ octahedra, given by ratio of the Ir-O(apical)/Ir-O(in-plane) distances, increases significantly with pressure.

Calculations of the Ir spin magnetic moments were also performed at 2.5 and 9.6 GPa. At 2.5 GPa, we found that the Ir spin moment is 0.048 $\mu_B$ in the {\it ab} plane, with a canting angle $\Theta=15^{\circ}$, very close to the octahedra tilt angle $\phi=14.4^{\circ}$ of the relaxed structure [see Figs. \ref{modes}(a) and \ref{DFT}(b)]. The near coincidence $\Theta \sim \phi$ is consistent with previous experiments \onlinecite{Ye,Boseggia} and calculations, being a signature of the $J_{eff}=1/2$ state \onlinecite{Kim,Jackeli,Liu}. The magnitude of the spin moment is consistent with a previous calculation at ambient pressure \onlinecite{Lado}. Using the compressed unit cell parameters at 9.6 GPa as input for our calculations, the calculated spin moment is washed out, indicating that the local Ir magnetic moment is unstable against a relatively small volumetric contraction ($\sim 5-10$ \%). This is expected for $5d$ systems, which are normally on the verge of magnetism. We should mention that the actual pressure value in which the Ir moment disappears in the calculations may depend on the choice of the on-site Coulomb $U$ and effective exchange parameters $J$, and is not further explored in this work. Experimentally, it was demonstrated that the ferromagnetic component of the Ir moment disappears at $\sim P_1$ \onlinecite{Haskel}, suggesting this is the critical pressure where the Ir ions become non-magnetic.

Density functional theory was also employed to perform {\it ab}-initio lattice dynamics calculations. A comparison with previous symmetry-resolved single crystal Raman studies \onlinecite{Cetin,Gretarsson,Gretarsson2} indicates that peak $M_3$ is a mode with $B_{2g}$ symmetry, peaks $M_2$ and $M_4$ are A$_{1g}$ modes and peak $M_1$ (hereby termed $M_1/M_1'$) is a superposition of an $A_{1g}$ and a $B_{2g}$ mode at nearby frequencies. Indeed, our calculations indicate $A_{1g}$ modes at 181, 260, and 588 cm$^{-1}$ and $B_{2g}$ modes at 173 and 371 cm$^{-1}$, in good agreement with the observed frequencies. The mechanical representations of such modes are given in Figs. \ref{modes}(c-g). The mode $M_1$ is a rotation of the IrO$_6$ octahedra along the $c$-axis combined with an in-phase Sr displacement along $c$, while $M_1^{'}$ is mostly an out-of-phase Sr vibration. $M_2$ is a pure rotation of the IrO$_6$ octahedra along $c$, $M_3$ is an in-plane bending of the IrO$_6$ octahedra, and $M_4$ is a stretching mode involving a modulation of the Ir-O(apical) distance. For completeness, the calculated mode frequencies and corresponding $\Gamma$-point mechanical representations of all Raman and infrared-active modes of Sr$_2$IrO$_4$ under the tetragonal space group $I$4$_1$/${acd}$ are given in Appendix A.

\section{Discussion}

\subsection{First-order structural phase transition}

An immediately visible result of our XRD experiment is the symmetry-lowering structural phase transition that occurs above 40 GPa, with a phase coexistence interval up to 55 GPa [see Fig. \ref{XRD_contour}]. The large collapse of the $c$-axis and expansion of one of the axes defining the {\it ab}-plane for the high pressure phase [see Figs. \ref{XRD_latticepar}(a) and \ref{XRD_latticepar}(b)] resembles the first order phase transition observed in Sr$_3$Ir$_2$O$_7$ at 54 GPa \onlinecite{Donnerer}. In that case, the XRD intensities could be well modeled when one of the perovskite bilayers of Sr$_3$Ir$_2$O$_7$ is translated in plane by half a unit cell. We attempted to perform similar rigid translations of Sr$_2$IrO$_4$ layers to model our XRD data, however without success. It is possible that significant intralayer atomic relaxation also occurs, so that the rigid layer translation model employed for Sr$_3$Ir$_2$O$_7$ \onlinecite{Donnerer} is not applicable for Sr$_2$IrO$_4$. Also, the Bragg peak overlap for the high-pressure phase of Sr$_2$IrO$_4$ is rather severe in our data [see Fig. \ref{XRD_Rietveld}(d)], which was detrimental to our attempts of  determining the crystal structure of the high-pressure phase. In any case, the similar critical pressures for this structural transition and comparable behavior of lattice parameters in the high pressure phases of Sr$_3$Ir$_2$O$_7$ and Sr$_2$IrO$_4$ argue in favor of a similar nature of the structural phase transitions of both materials. It is interesting to note that, for the case of Sr$_3$Ir$_2$O$_7$, the high-pressure structural phase transition is followed by a metallization of the material \onlinecite{Ding}. It is possible that a similar effect occurs for Sr$_2$IrO$_4$. Resistivity measurements above $\sim 55$ GPa are necessary to confirm or dismiss this expectation.

\subsection{Phonon and lattice anomalies below 40 GPa}

Besides the structural phase transition observed at high pressures, marked anomalies of lattice parameters, strain, and phonon Raman spectra were found at lower pressures, i.e., well within the tetragonal phase. Particularly clear anomalies occur at $P_1=17$ GPa. Above this pressure, the volume compressibility is significantly reduced [Fig. \ref{XRD_latticepar}(c)] and the $c/a$ ratio stabilizes at $\sim 4.74$ [Fig. \ref{XRD_c_over_a}]. Also the Raman-active phonon modes $M_1$ and $M_2$ weaken sensibly with low pressures and disappear above $P_1$ [Figs. \ref{spectra}, \ref{fitresults}(i) and \ref{fitresults}(j)]. Phonon $M_3$ shows an asymmetric lineshape [Fig. \ref{fitresults}(m) and Appendix D] and hardens anomalously above $P_1$. Finally, mode $M_4$ weakens above $\sim P_1$. It is interesting to note that a previous report indicates that the ferromagnetic component measured by x-ray magnetic circular dichroism disappears above $\sim P_1$ \onlinecite{Haskel}. This reference pressure also marks a change of behavior of transport properties of Sr$_2$IrO$_4$; for instance, the electrical resistance for the $a$-axis at 50 K shows a three orders of magnitude reduction from ambient pressure to $P_1$, stabilizing at a nearly constant value between $P_1$ and $\sim 30$ GPa \onlinecite{Haskel}. These combined results point to a phase transition at $P_1$.

Our discussion on the nature of the phase transition at $P_1$ begins by considering the possibility of a subtle structural phase transition, perhaps not directly captured by our XRD data. Such transition could either increase or reduce the symmetry of the crystal lattice. A structural symmetrization would occur if the IrO$_6$ octahedral tilt angle $\phi$ [Fig. \ref{modes}(a)] went to zero at $P_1$, leading to a phase transition from the $I4_1/acd$ to the $I4/mmm$ space group. Although this hypothetical symmetrization with pressure is counter intuitive and is not supported by our DFT results [see Fig. \ref{DFT}(b)], it might in principle explain the disappearance of the ferromagnetic moment at $\sim P_1$ observed by x-ray magnetic circular dichroism \onlinecite{Haskel}, as the magnetic canting angle $\Theta$ is known to be coupled with $\phi$. To test this possibility, the XRD profile at 18.1 GPa [Fig. \ref{XRD_Rietveld}(b)] was analyzed under both $I4_1/acd$ and $I4/mmm$ symmetries. All the observed Bragg reflections within our sensitivity are allowed for both space groups, however, the Rietveld fit shown in Fig. \ref{XRD_Rietveld}(b) for the $I4_1/acd$ space group had much improved $R$-factors with respect to the $I4/mmm$ symmetry (for instance, $R_{wp}=3.73$ \% for $I4_1/acd$ and $R_{wp}=4.88$ \% for $I4/mmm$), indicating that the $I4/mmm$ high-symmetry structure is not adequate to describe the phase above $P_1$. The same conclusion is reached by an analysis of our Raman scattering results, as the observation of the in-plane oxygen bending $M_3$ mode [see Fig. \ref{spectra}] shows that such oxygen is not at an inversion center either below or above $P_1$, dismissing a transition to the $I4/mmm$ high-symmetry structure.

In opposition to the structural symmetrization hypothesis discussed above, we also consider the possibility of a symmetry-reduction structural transition at $P_1$ that might have not been immediately noticed in our XRD profiles due to limited resolution. Indeed, the Bragg peak broadening with increasing pressures [see, for instance, Fig. \ref{XRD_Rietveld}(e)] reduces the resolution of this experiment to capture subtle structural phase transitions. A complicating factor is the reported effects of non-hydrostaticity of our pressure transmitting medium (Neon) above 15 GPa \onlinecite{Klotz}, since this threshold pressure is very similar to $P_1$. In this way, it is not clear whether the sudden increment of the Gaussian strain broadening parameter $U$ between 15 and 20 GPa [see Fig. \ref{XRD_strain}(a)] is due to loss of pressure hydrostaticity or rather an intrinsic manifestation of the phase transition at $P_1$. On the other hand, the significantly wider 200 Bragg peak with respect to the nearby 116 peak [see Fig. \ref{XRD_Rietveld}(e)], and, more generally, the dominant $S_{400}$ term over the other Lorentzian strain broadening parameters between $P_1$ and $P_2$ [see Fig. \ref{XRD_strain}(b)] is not trivially explained in terms of non-hydrostatic pressure conditions. Such anisotropic broadening rather indicates a four-fold symmetry-breaking strain in this pressure region caused by a small or non-cooperative orthorhombic distortion of the crystal lattice.


Additional information about the transition at $P_1$ is gained by a careful consideration of our Raman scattering data. First of all, the vanishing Raman cross sections of modes $M_1$ and $M_2$ with increasing pressures up to $P_1$ [see Figs. \ref{spectra}, \ref{fitresults}(i), and  \ref{fitresults}(j)] is not a direct consequence of symmetry reduction suggested by XRD, since the opposite trend, i.e., additional Raman modes above $P_1$, would be naively expected. 
The apparent disappearance of the peaks $M_1$ and $M_2$ as $P \rightarrow P_1$ must be rather ascribed to a large reduction of the corresponding Raman tensor elements, which is a direct manifestation of significant changes in the electronic structure at $P_1$.
In addition, the broadening of the modes $M_3$ and $M_4$ above $P_1$ [see Figs. \ref{fitresults}(g) and \ref{fitresults}(h)] shows clear correspondence with the evolution of the $S_{400}$ Lorentzian Bragg peak broadening term, indicating that the symmetry-breaking lattice strain inferred above causes relevant impact in the phonon lifetimes.

Our DFT calculations provide yet another hint on the nature of the phase transition at $P_1$ by indicating that the local Ir magnetic moment is unstable against application of pressure. This result, allied to the disappearance of the x-ray magnetic circular dichroism signal above $P_1$ \onlinecite{Haskel}, indicate that the state above $P_1$ is actually non-magnetic.

Armed with the above considerations, we now proceed to discuss the possible nature of the phase above $P_1$. We enumerate the possible scenarios:

(i) {\it Charge-density wave instability}: the quenching of the local Ir magnetic moments at $P_1$ and the proximity to a metal-insulator transition may favor an emerging charge-density wave phase. This scenario is able to explain quite naturally the symmetry-breaking lattice strain and reduced phonon lifetimes observed here. In addition, it is consistent with the avoided metallization above $P_1$ \onlinecite{Caorev,Haskel}, since a CDW state presumably restricts the conduction channels. It also reinforces the parallel with cuprate superconductors, once several observations of CDW instability have been reported for cuprates \onlinecite{Wu,Chang,Ghiringhelli,Kawasaki}.

(ii) {\it Nematic instability}: the Ir $5d$ electrons may show a tendency to present uneven proportions of $xz$ and $yz$ orbitals  above $P_1$, leading to an electronic nematic instability. In this case, a parallel could be traced with respect to the Fe-based parent superconductors \onlinecite{Fernandes}. However, such hypothetical nematic state would be associated with non-magnetic Ir ions ($P > P_1$), while in the parent Fe-based superconductors the nematic state is actually favored by magnetism \onlinecite{Fernandes}. We should mention that this scenario offers little insight into the avoided metallization of Sr$_2$IrO$_4$ above $P_1$.

(iii) {\it IrO$_6$ octahedral rotations}: in the crystal structure of Sr$_2$IrO$_4$ at ambient conditions, the IrO$_6$ octahedra are rotated along the $c$-axis only [see Fig. \ref{modes}(a)]. It is possible that external pressures above $P_1$ favor an additional rotation of the octahedra along the tetragonal $a$ (or $b$) axis, increasing the compactness of the structure and favoring the compression along the $c$-direction. Such atomic rearrangement, if cooperative, would lead to a symmetry reduction with respect to the tetragonal unit cell, leading to a structural phase transition that was not observed. However, non-cooperative or short-range-ordered octahedral tilts are also possible and would lead to anisotropic strain and selective Bragg peak broadening, in accordance to our observations. This scenario is consistent with both XRD and Raman data. In fact, the stabilization of the $c/a$ ratio between $P_1$ and $P_2$ [see Fig. \ref{XRD_c_over_a}] is consistent with a more isotropic compression of the unit cell that would occur if the IrO$_6$ octahedra were allowed to rotate along more than one axis. Also, the structural disorder above $P_1$ in this scenario might induce Anderson localization and possibly explain the avoided metallization above $P_1$. However, this scenario provides little insight into the loss of the Ir magnetic moments above $P_1$ \onlinecite{Haskel}.

(iv) {\it isospin-flop transition}: in the phase diagram of iridates, the non-colinear magnetic ground state competes with a colinear state where the isospins are oriented along $c$ \onlinecite{Jackeli,Lado}. The colinear state is favored by increasing tetragonal distortion of the IrO$_6$ octahedra. Considering that both the $c/a$ ratio [Fig. \ref{XRD_c_over_a}] and octahedral elongation [Fig. \ref{DFT}(c)] increase with pressure, it is possible that a pressure-induced isospin-flop transition takes place at $P_1$ with a possible loss of the spin-orbit entanglement characteristic of the $J_{eff}=1/2$ state \onlinecite{Lado}. Such change of electronic configuration is consistent with the anomalous hardening of mode $M_3$ above $P_1$ [see Fig. \ref{fitresults}(c)] and the loss of x-ray magnetic circular dychroism signal \onlinecite{Haskel}. On the other hand, this scenario is not able to explain the fourfold symmetry-breaking instability observed above $P_1$. Also, it is not clear whether manifestations of such isospin-flop transition could be observed at 300 K, much above the magnetic ordering temperature of Sr$_2$IrO$_4$, $T_N=240$ K. This scenario also presumes that the local Ir magnetic moments survive to pressures above $P_1$, which is questionable considering the instability of such moments through volume contraction suggested by our density functional theory calculations.

It is interesting to note that scenarios (i) and (ii) above imply a ``competing order'' to superconductivity that is strong enough to be manifested in our measurements at room temperature, at least for undoped Sr$_2$IrO$_4$. Thus, the identification of the correct scenario for the phase transition at $P_1$ may shed light onto the fundamental reasons for the absence of the widely expected superconductivity of Sr$_2$IrO$_4$ under electron doping. While we consider that a charge density wave instability is the most physically sound hypothesis to explain all the presently available data and push forward the similarity with the cuprates, the other possibilities enumarated above cannot be completely excluded, and further experiments are necessary to settle this issue.

It is evident from our experimental data that yet another change of regime takes place at $P_2=30$ GPa. In fact, the $c/a$ ratio increases [Fig. \ref{XRD_c_over_a}], mode $M_1$ reappears [Fig. \ref{fitresults}(i)] and mode $M_4$ is washed out [Fig. \ref{fitresults}(l)] above $P_2$. Also, the abrupt rise of the $S_{220}$ Lorentzian broadening parameter above $P_2$ [Fig. \ref{XRD_strain}(b)] indicates that the four-fold symmetry-breaking lattice strain or instability between $P_1$ and $P_2$ gives place to a more isotropic structural disorder above $P_2$. Additional changes are seen at $P_3=40$ GPa, most notably a resymmetrization of the $M_3$ lineshape [Fig. \ref{fitresults}(m) and Appendix D] and a further elongation of the tetragonal crystal structure [Fig. \ref{XRD_c_over_a}]. Despite the proximity of $P_3$ with the pressure range where the first-order structural phase transition occurs (40-55 GPa), the changes at $P_3$ enumarated above do not seen to be associated with the high-pressure monoclinic phase. For instance, the monoclinic phase shows a colapse of the $c$-axis, while the tetragonal phase shows an increased $c/a$ ratio above $P_3$. Also, the monoclinic phase does not show visible Raman bands [Fig. \ref{spectra}], thus the observed phonon anomalies are most likely associated with relevant changes within the tetragonal phase.

Considering that the state of Sr$_2$IrO$_4$  between $P_1$ and $P_2$ was not unambiguously determined, an attempt to identify the phase beyond $P_2$ with the data presently at hands could not be carried out without a large dose of speculation. However, some important constraints do apply. Besides our XRD and Raman data, previous x-ray absorption data show a significant reduction in the Ir $L_3/L_2$-edges threshold ratio between 30 and 40 GPa, also evidencing changes of the Ir electronic state. Considering the increasing tetragonal $c/a$ ratio in this pressure range, it is reasonable to suppose that a Jahn-Teller distortion of the IrO$_6$ octahedra driven by the Ir $t_{2g}$ electrons is an important ingredient to understand the physical behavior of this material, at least at $P>P_2$.

\section{Conclusions}

In conclusion, we performed a comprehensive XRD, Raman scattering and DFT study of the effect of external pressure on the crystal lattice and vibrations of Sr$_2$IrO$_4$ at room temperature.  Lattice dynamics calculations at ambient pressure show good agreement with experimental phonon frequencies, allowing for the identification of the corresponding normal modes of vibration. A first-order structural phase transition from tetragonal to monoclinic phases is observed at $\sim 50$ GPa with a broad phase coexistence range $40 < P < 55$ GPa, characterized by a collapse of the tetragonal $c$-axis and an expansion of one of the in-plane lattice parameters in the high-pressure phase, in close similarity with the related Ruddlesden-Popper iridate Sr$_3$Ir$_2$O$_7$ \onlinecite{Donnerer}. Remarkably, a number of lattice and phonon anomalies were observed within the tetragonal phase at 17, 30, and 40 GPa, which were ascribed to crossovers between competing electronic states. A number of possible alternative scenarios were considered and critically discussed for the transition at 17 GPa. The states above 30 GPa were found to show an increased elongation of the crystal structure along the $c$-axis, suggesting an active role of the Jahn-Teller effect in iridate physics despite the strong spin-orbit coupling.

\begin{appendix}

\renewcommand{\thefigure}{A\arabic{figure}}
\renewcommand{\thetable}{A\arabic{table}}
\setcounter{figure}{0}  

\section{Phonon symmetry analysis and lattice dynamics calculations}

\begin{figure*}
	\includegraphics[width=1.0 \textwidth]{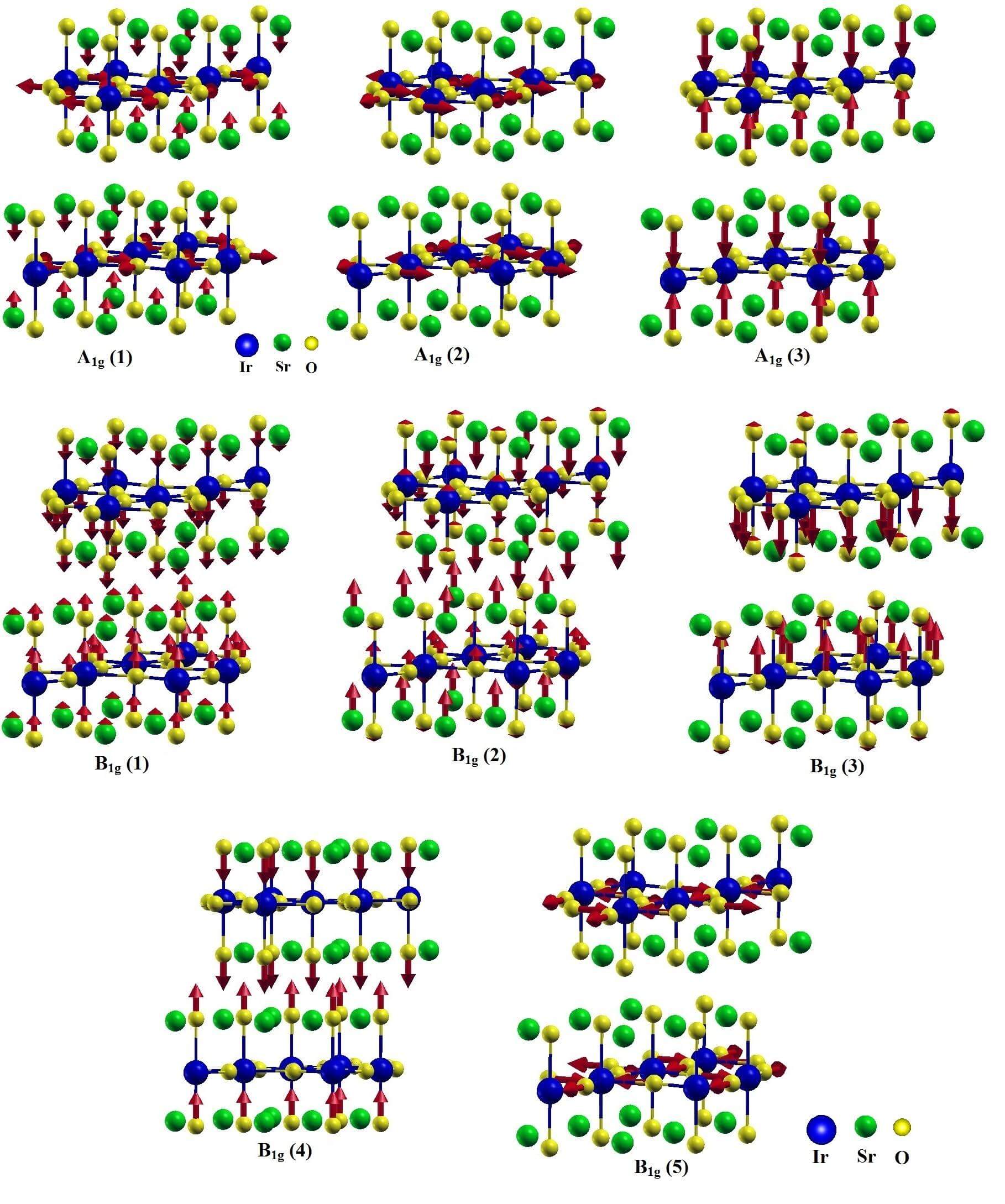}
	\caption{\label{Modes1} Mechanical representations of Raman-active $A_{1g}$ and $B_{1g}$ modes (see also Table \ref{freqRaman}).}
\end{figure*}

\begin{figure*}
	\includegraphics[width=1.0 \textwidth]{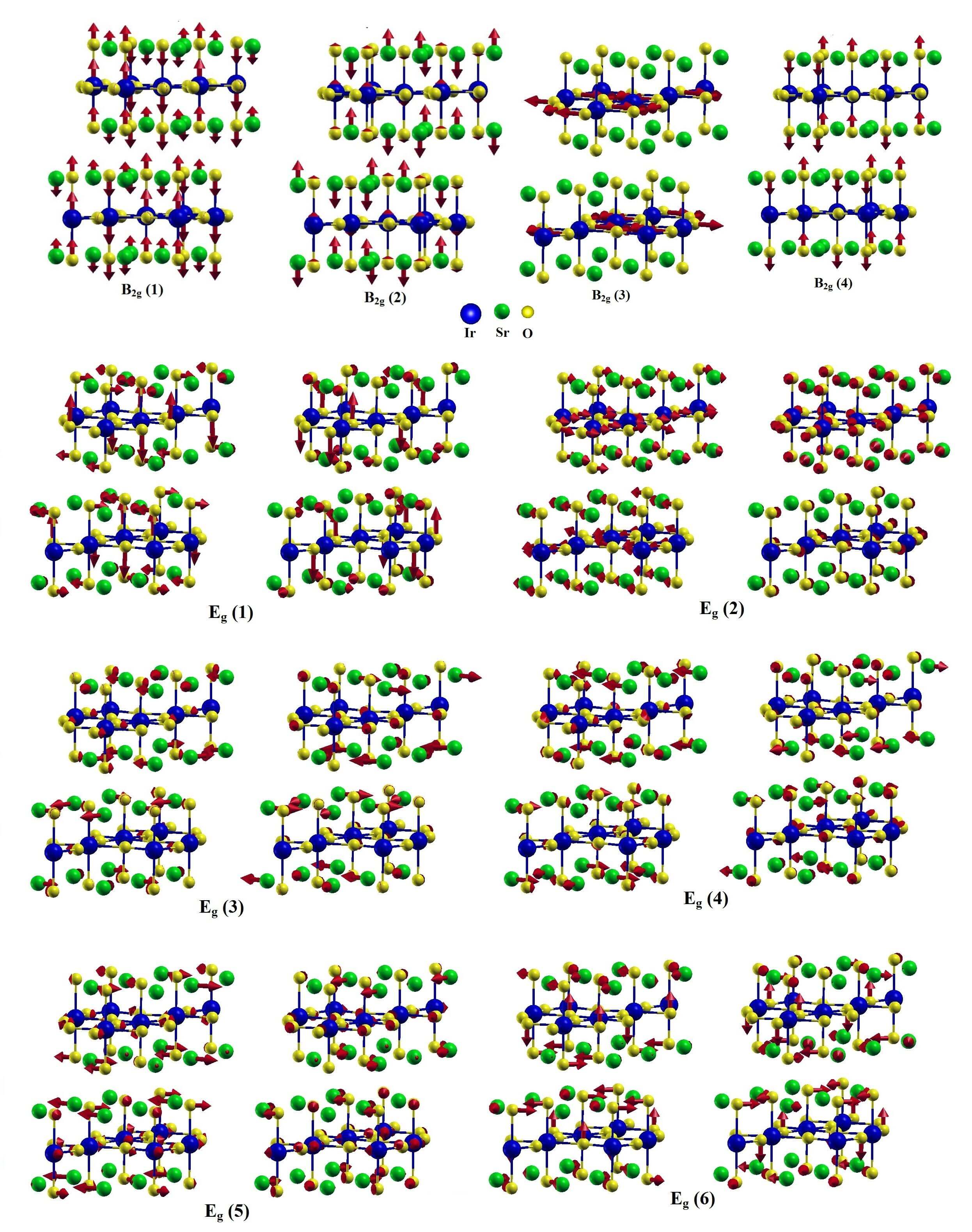}
	\caption{\label{Modes2} Mechanical representations of Raman-active $B_{2g}$ and $E_{g}(1)$-$E_{g}(6)$ modes (see also Table \ref{freqRaman}).}
\end{figure*}

\begin{figure*}
	\includegraphics[width=1.0 \textwidth]{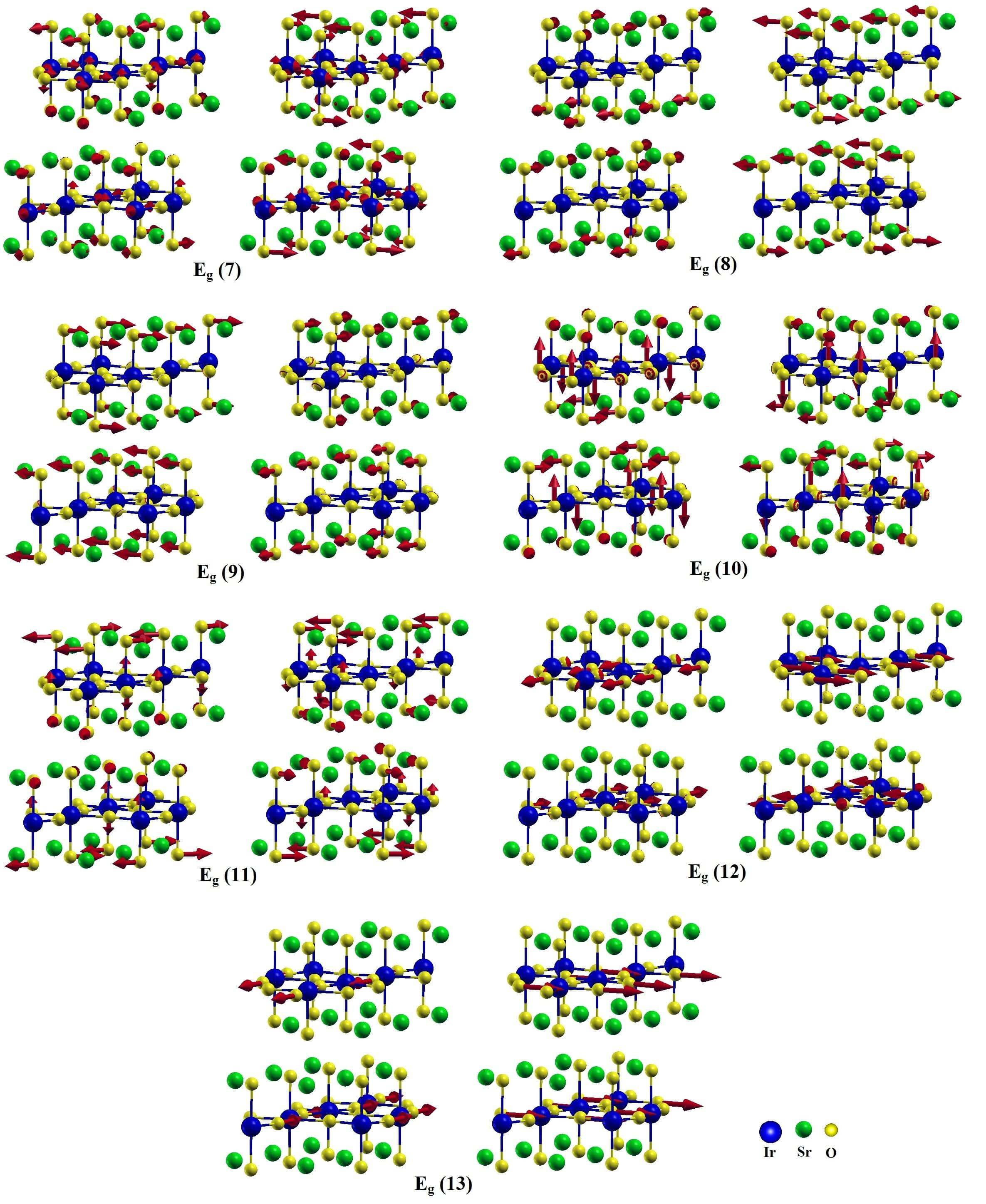}
	\caption{\label{Modes3} Mechanical representations of Raman-active $E_{g}(7)$-$E_{g}(13)$ modes (see also Table \ref{freqRaman}).}
\end{figure*}

\begin{figure*}
	\includegraphics[width=1.0 \textwidth]{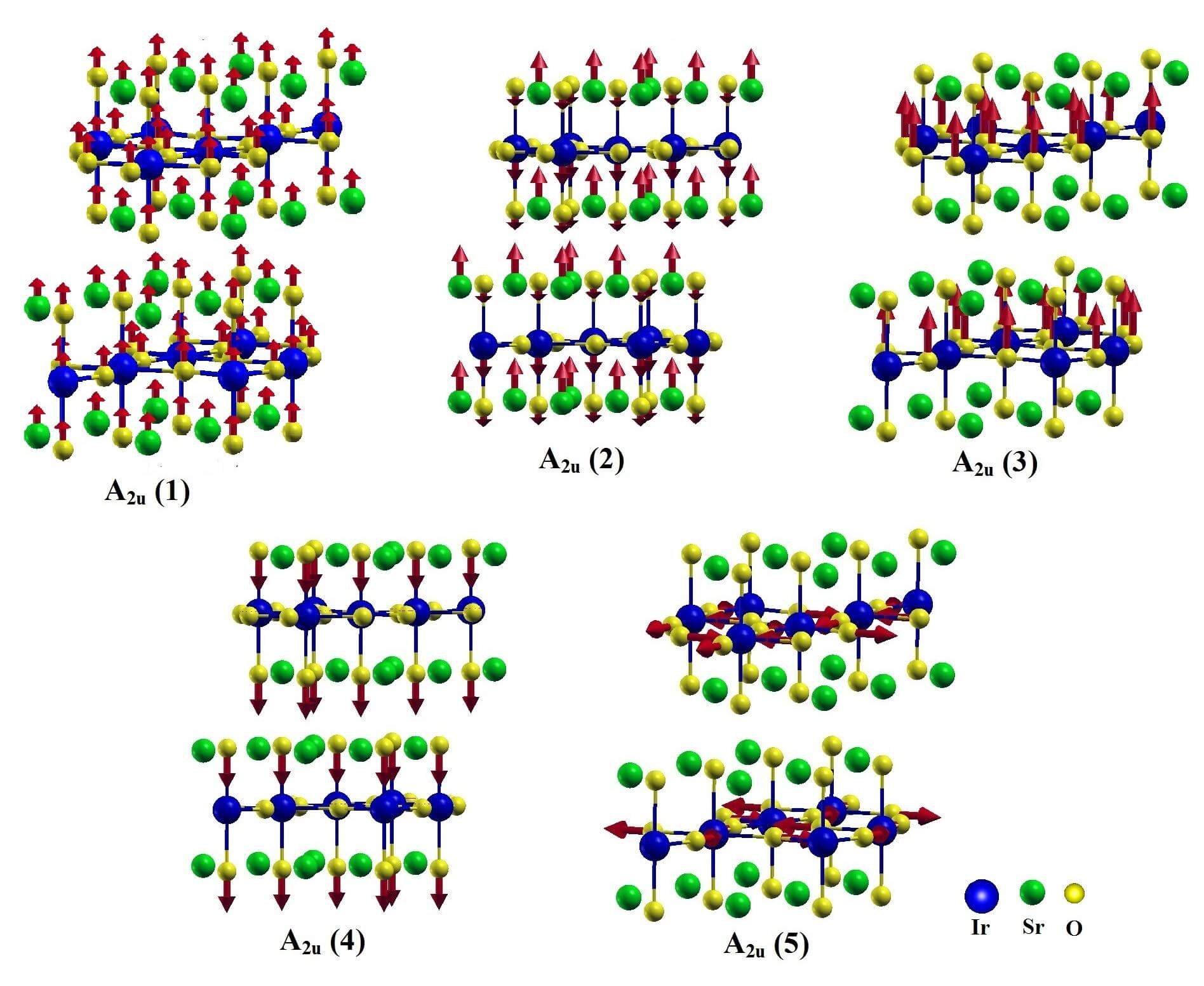}
	\caption{\label{Modes4} Mechanical representations of the infrared-active $A_{2u}$ modes (see also Table \ref{freqIR}).}
\end{figure*}

\begin{figure*}
	\includegraphics[width=1.0 \textwidth]{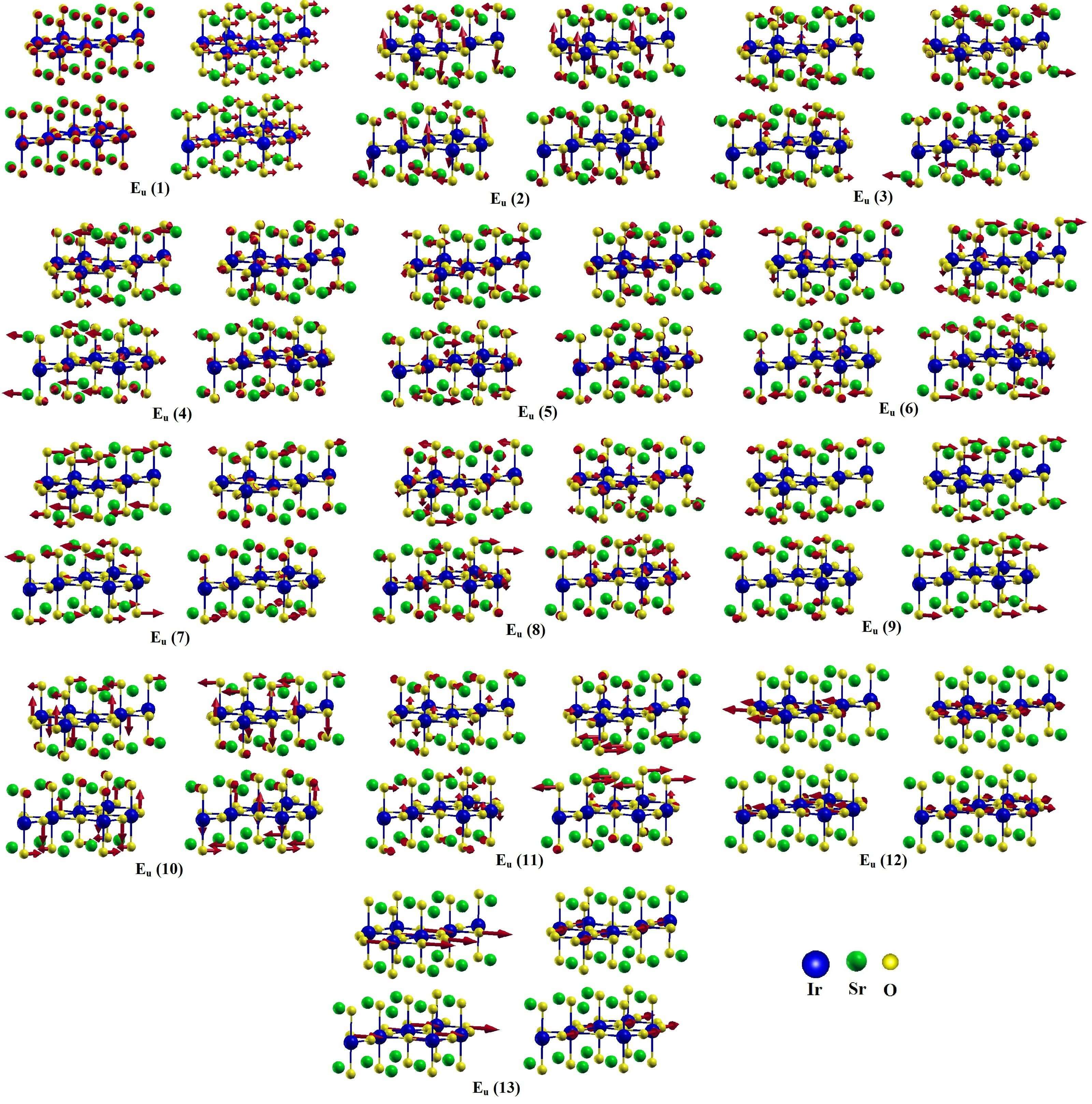}
	\caption{\label{Modes5}  Mechanical representations of the infrared-active $E_{u}$ modes (see also Table \ref{freqIR}).}
\end{figure*}

The conventional body-centered tetragonal cell of Sr$_2$IrO$_4$ contains eight formula units, totaling 56 atoms, thus the primitive cell contains 28 atoms and 84 vibrational degrees of freedom. The symmetry analysis of these vibrational modes is given in Table \ref{ModesTet}. The {\it ab}-initio calculated frequencies of the 25 Raman-active and 18 infrared-active modes are given in Tables \ref{freqRaman} and \ref{freqIR}, respectively, showing good agreement with experimental data. Imaginary calculated frequencies were obtained for the $E_g(1)$ and $E_u(2)$ modes, which is a spurious result associated with our choice to use the experimental (unrelaxed) crystal structure \onlinecite{Crawford} as input for the lattice dynamics calculations. The corresponding mechanical representations are displayed in Figs. \ref{Modes1}-\ref{Modes3} and \ref{Modes4}-\ref{Modes5}, respectively.

\begin{table}
	\caption{\label{ModesTet} Wyckoff positions and irreducible representations of $\Gamma$-point phonon modes for the tetragonal phase of Sr$_2$IrO$_4$, space group $I4_1/acd$ ($D_{4h}^{20}$, No. 142). The corresponding Raman tensors are also given.}
	
	\begin{ruledtabular}
		\begin{tabular}{ccc}
			Atom & Wyckoff & $\Gamma$ -point phonon modes \\
			& position & \\
			\hline
			Sr & $16d$ & $A_{1u}$+$A_{2u}$+$A_{1g}$+$A_{2g}$+$B_{1u}$+$B_{2u}$+$B_{1g}$+\\
			& & +$B_{2g}$+4$E_{u}$+4$E_g$ \\
			Ir & $8a$ & $A_{1u}$+$A_{2u}$+$B_{1g}$+$B_{2g}$+2$E_u$+2$E_g$  \\
			O(1) & $16d$ & $A_{1u}$+$A_{2u}$+$A_{1g}$+$A_{2g}$+$B_{1u}$+$B_{2u}$+$B_{1g}$+\\
			& & +$B_{2g}$+4$E_{u}$+4$E_g$ \\
			O(2) & $16f$ & $A_{1g}$+$A_{1u}$+2$A_{2u}$+2$A_{2g}$+2$B_{1g}$+2$B_{1u}$+$B_{2g}$+ \\
			& & +$B_{2u}$+3$E_g$+3$E_u$ \\
			\hline
			\multicolumn{3}{c}{Classification:}  \\
			\multicolumn{3}{c}{$\Gamma_{Raman}$=3$A_{1g}$+5$B_{1g}$+4$B_{2g}$+13$E_g$} \\
			\multicolumn{3}{c}{$\Gamma_{IR}$=5$A_{2u}$+13$E_u$ ($\Gamma_{Acoustic}$=$A_{2u}$+$E_u$)} \\
			\multicolumn{3}{c} {$\Gamma_{Silent}$=4A$_{1u}$+4A$_{2g}$+4$B_{1u}$+3$B_{2u}$ } \\
			\hline
			\multicolumn{3}{c}{Raman tensors:} \\
			\multicolumn{3}{c}{ $A_{1g} \rightarrow
				\begin{pmatrix}
				a & 0 & 0 \\
				0 & a & 0 \\
				0 & 0 & b \\ 
				\end{pmatrix}$,
				$B_{1g} \rightarrow
				\begin{pmatrix}
				c & 0 & 0 \\
				0 & -c & 0 \\
				0 & 0 & 0 \\ 
				\end{pmatrix},
				B_{2g} \rightarrow
				\begin{pmatrix}
				0 & d & 0 \\
				d & 0 & 0 \\
				0 & 0 & 0 \\ 
				\end{pmatrix}$} \\
			\multicolumn{3}{c}{$E_{g} \rightarrow
				\begin{pmatrix}
				0 & 0 & 0 \\
				0 & 0 & e \\
				0 & e & 0 \\ 
				\end{pmatrix},
				\begin{pmatrix}
				0 & 0 & -e \\
				0 & 0 & 0 \\
				-e & 0 & 0 \\ 
				\end{pmatrix}$} \\
			
		\end{tabular}
	\end{ruledtabular}
\end{table}

\begin{table}
	\caption{\label{freqRaman} First-principle-calculated frequencies of the Raman-active modes represented in Figs. \ref{Modes1}, \ref{Modes2}, and \ref{Modes3}, and comparison with single-crystal experimental data.}
	
	\begin{ruledtabular}
		\begin{tabular}{ccc}
Mode & Calc (cm$^{-1}$) & Observ. (cm$^{-1}$) \\
& & Ref. \onlinecite{Gretarsson2} \\
\hline
$A_{1g}(1)$ & 181 & 188 \\
$A_{1g}(2)$ & 260 & 278 \\
$A_{1g}(3)$ & 588 & 562 \\
$B_{1g}(1)$ & 114 & \\
$B_{1g}(2)$ & 167 & \\
$B_{1g}(3)$ & 359 & \\
$B_{1g}(4)$ & 596 & \\
$B_{1g}(5)$ & 808 & $\sim 735$ \\
$B_{2g}(1)$ & 134 & \\
$B_{2g}(2)$ & 173 & \\
$B_{2g}(3)$ & 371 & 395 \\
$B_{2g}(4)$ & 513 & \\
$E_{g}(1)$ & imag. & \\
$E_{g}(2)$ &  53 & \\
$E_{g}(3)$ & 91 & \\
$E_{g}(4)$ & 108 & \\
$E_{g}(5)$ & 132 & \\
$E_{g}(6)$ & 186 & 191 \\
$E_{g}(7)$ & 203 & \\
$E_{g}(8)$ & 205 & \\
$E_{g}(9)$ & 266 & \\
$E_{g}(10)$ & 291 & \\
$E_{g}(11)$ & 314 & \\
$E_{g}(12)$ & 380 & \\
$E_{g}(13)$ & 751 & \\
		\end{tabular}
	\end{ruledtabular}
\end{table}

\begin{table}
	\caption{\label{freqIR} Calculated frequencies of the infrared-active modes represented in Figs. \ref{Modes4} and \ref{Modes5}, and comparison with experimental data.}
	
	\begin{ruledtabular}
		\begin{tabular}{ccc}
Mode & Calc (cm$^{-1}$) & Observ. (cm$^{-1}$) \\
& & Ref. \onlinecite{Propper} \\
\hline
$A_{2u}(1)$ & acoustic & \\
$A_{2u}(2)$ & 172 & 192 \\
$A_{2u}(3)$ & 340 & 373 \\
$A_{2u}(4)$ & 502 & 515 \\
$A_{2u}(5)$ & 808  & \\
$E_{u}(1)$ & acoustic & \\
$E_{u}(2)$ & imag. & \\
$E_{u}(3)$ & 84 & 103 \\
$E_{u}(4)$ & 101 & 115 \\
$E_{u}(5)$ & 120 & 138 \\
$E_{u}(6)$ & 174 & \\
$E_{u}(7)$ & 194 & \\
$E_{u}(8)$ & 204 & 214 \\
$E_{u}(9)$ & 257 & 270 \\
$E_{u}(10)$ & 293 & 282.5 \\
$E_{u}(11)$ & 314 & 324 \\
$E_{u}(12)$ & 381 & 367 \\
$E_{u}(13)$ & 751 & 664 \\
		\end{tabular}
	\end{ruledtabular}
\end{table}

\section{{\it dc}-Magnetization}

\begin{figure}
	\includegraphics[width=0.5\textwidth]{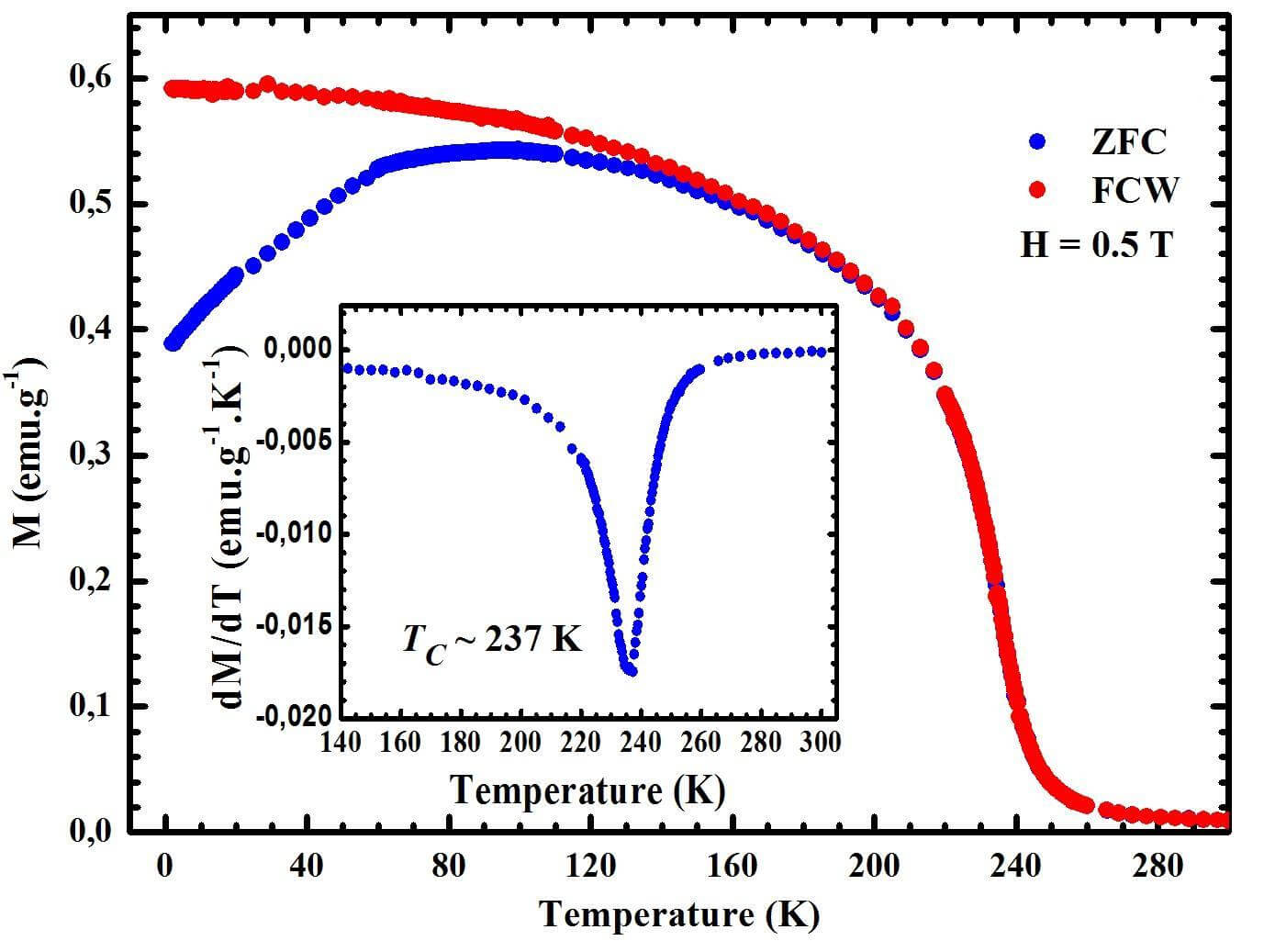}
	\begin{quotation}
		\caption{{\it dc}-magnetization taken with $H = 0.5$ T under warming after zero field cooling (ZFC, blue symbols) and warming after field cooling (FCW, red symbols). The inset shows the ZFC  magnetization derivative curve, marking the magnetic transition temperature TC = 237 K.}
		\label{Mag}
	\end{quotation}
\end{figure}

{\it dc}-magnetization measurements as a function of temperature at ambient pressure were performed for the polycrystalline sample employed in this work (see Fig. \ref{Mag}). The experiment was performed with a commercial Superconducting Quantum Interference Device (SQUID) magnetometer. The measurements taken under warming after field cooling (FCW, $H=0.5$ T) show a monotonic reduction of magnetization on warming, marking a magnetic ordering transition temperature of $T_C=237$ K (see inset of Fig. \ref{Mag}). The magnetization taken under warming after zero field cooling (ZFC) is smaller than the FCW magnetization up to $\sim 160$ K. The difference between the FCW and ZFC curves may be due to pinning of domain walls caused by lattice deffects. 

\section{Supplementary x-ray diffraction data} 

\begin{figure}
	\includegraphics[width=0.45\textwidth]{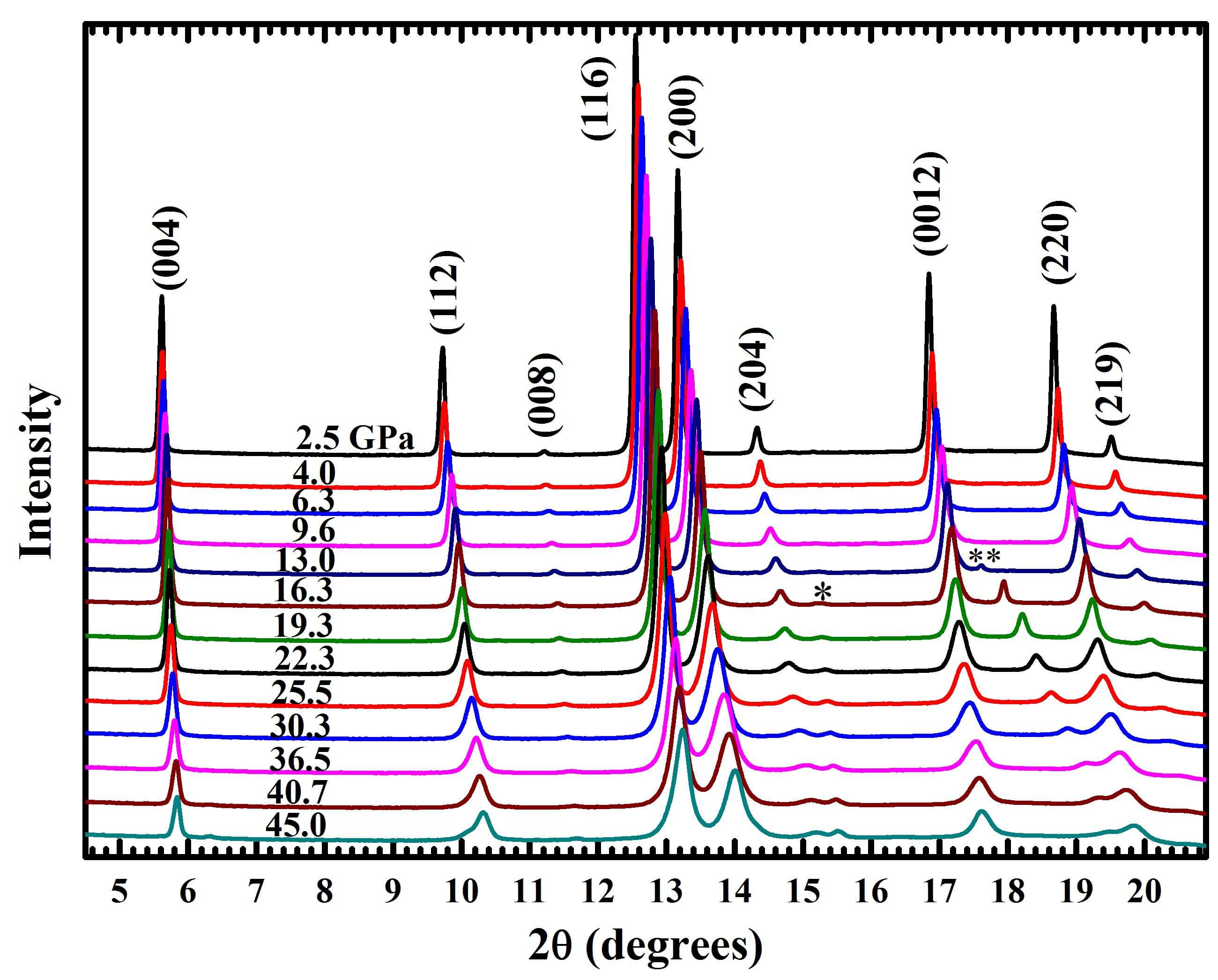}
	\begin{quotation}
		\caption{Pressure-dependent x-ray diffraction up to 45 GPa for the preliminary run. The peaks marked as '*' and '**' represent the diffraction peaks from Rhenium gasket and Neon, respectively.}
		\label{XRD_run1}
	\end{quotation}
\end{figure}

We collected a preliminary set of XRD and Raman scattering data that are consistent with data shown in the main text. The raw XRD data for this preliminary study are shown in Fig. \ref{XRD_run1} [see also Figs. \ref{XRD_latticepar}(a-b) of the main paper for a comparison of refined lattice parameters for both runs]. We should mention that the maximum pressure in this case was only 45 GPa. Thus, the first-order structural transition was not observed, except for incipient features such as a shoulder in the 112 reflection and a small peak at 6.2 $^{\circ}$ at 45 GPa that is due to the 004 reflection of the minority monoclinic phase in the phase coexistence regime.

\section{Supplementary Raman scattering data}

\begin{figure}
	\includegraphics[width=0.5\textwidth]{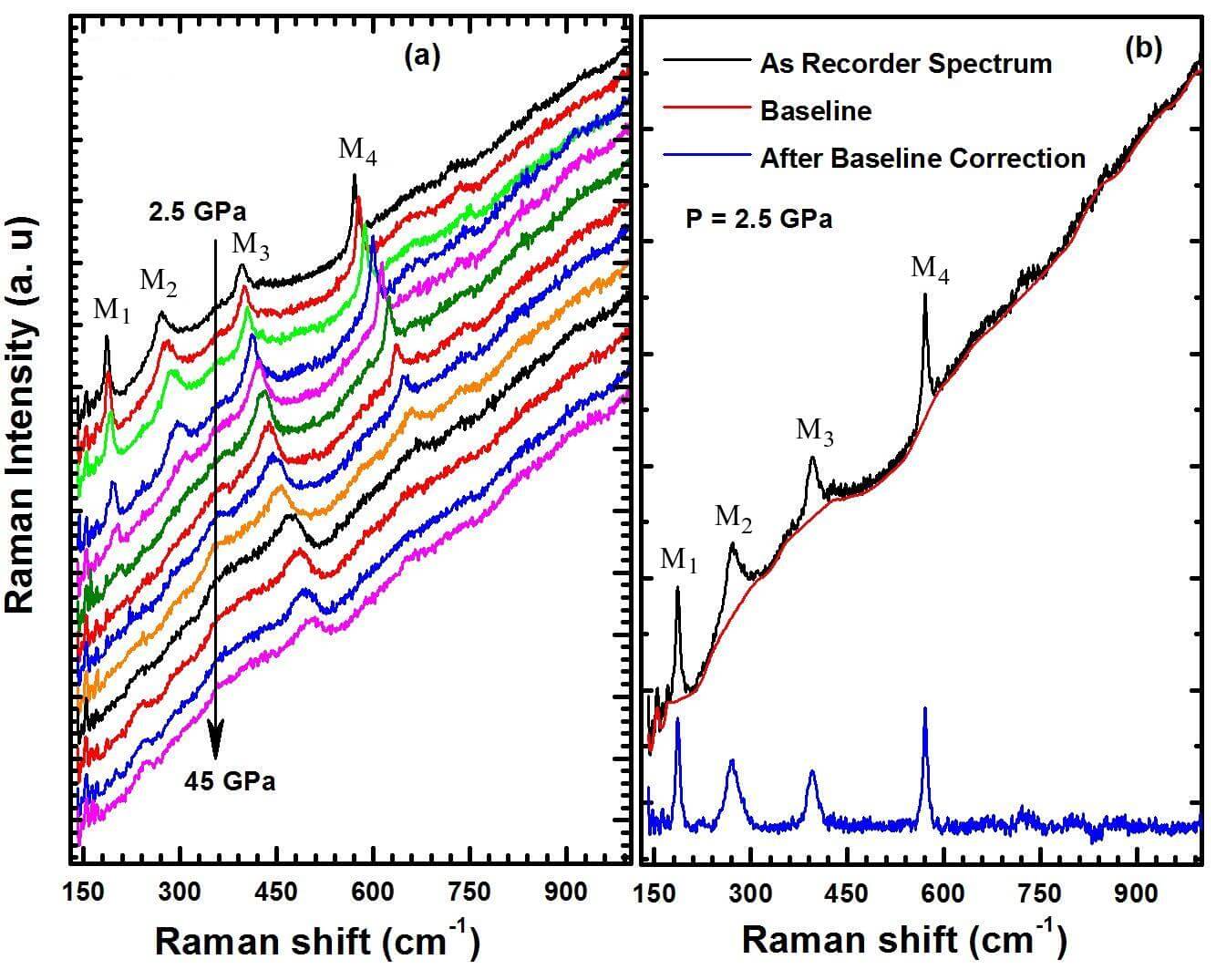}
	\begin{quotation}
		\caption{(a) Pressure dependent Raman spectra for the preliminary run. (b) Baseline correction at $P = 2.5$ GPa. The red solid line in (b) represents the baseline of the spectrum and the blue line is the spectrum after baseline correction.}
		\label{Raman_background}
	\end{quotation}
\end{figure}



\begin{figure}
	\includegraphics[width=0.5\textwidth]{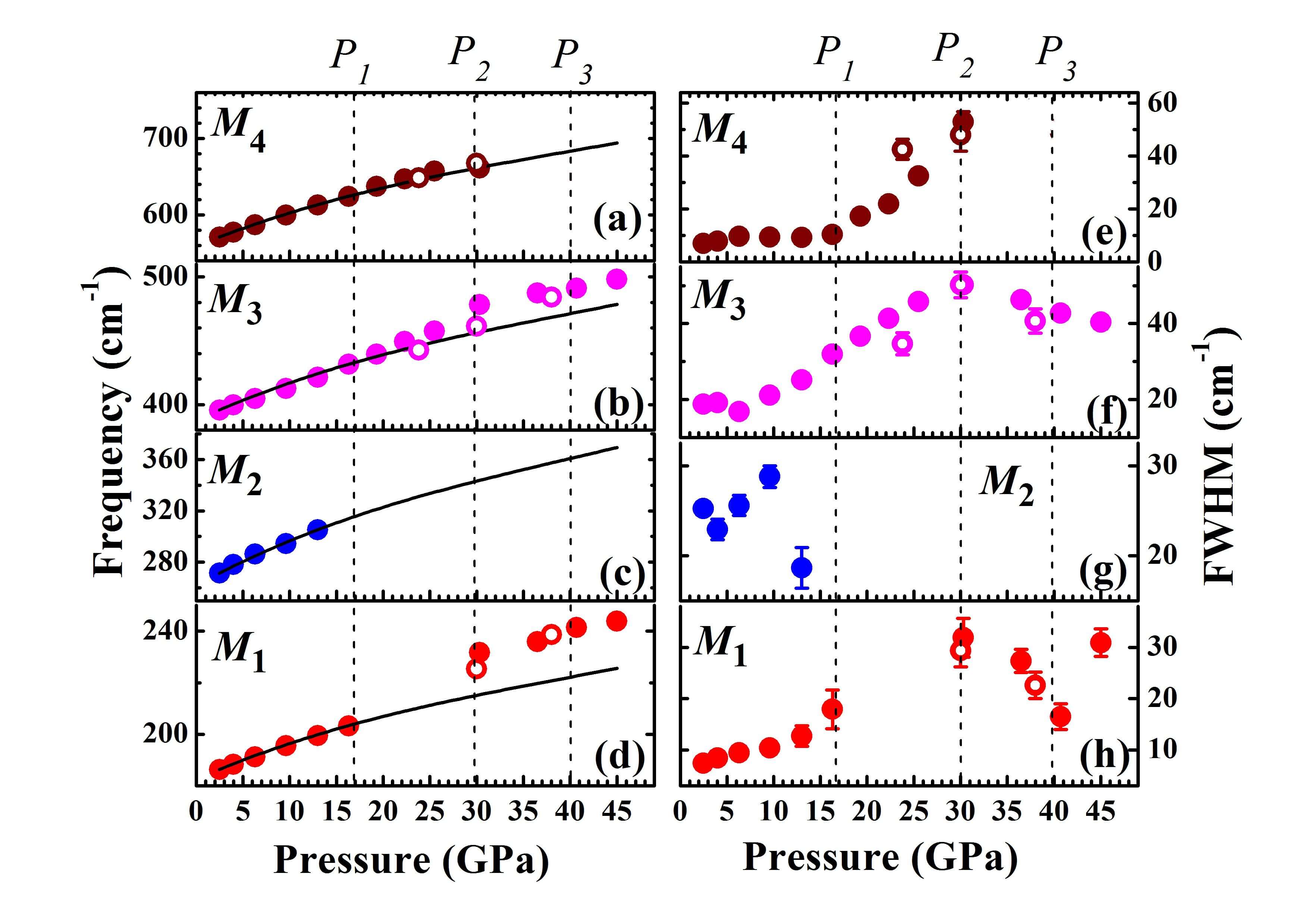}
	\includegraphics[width=0.5\textwidth]{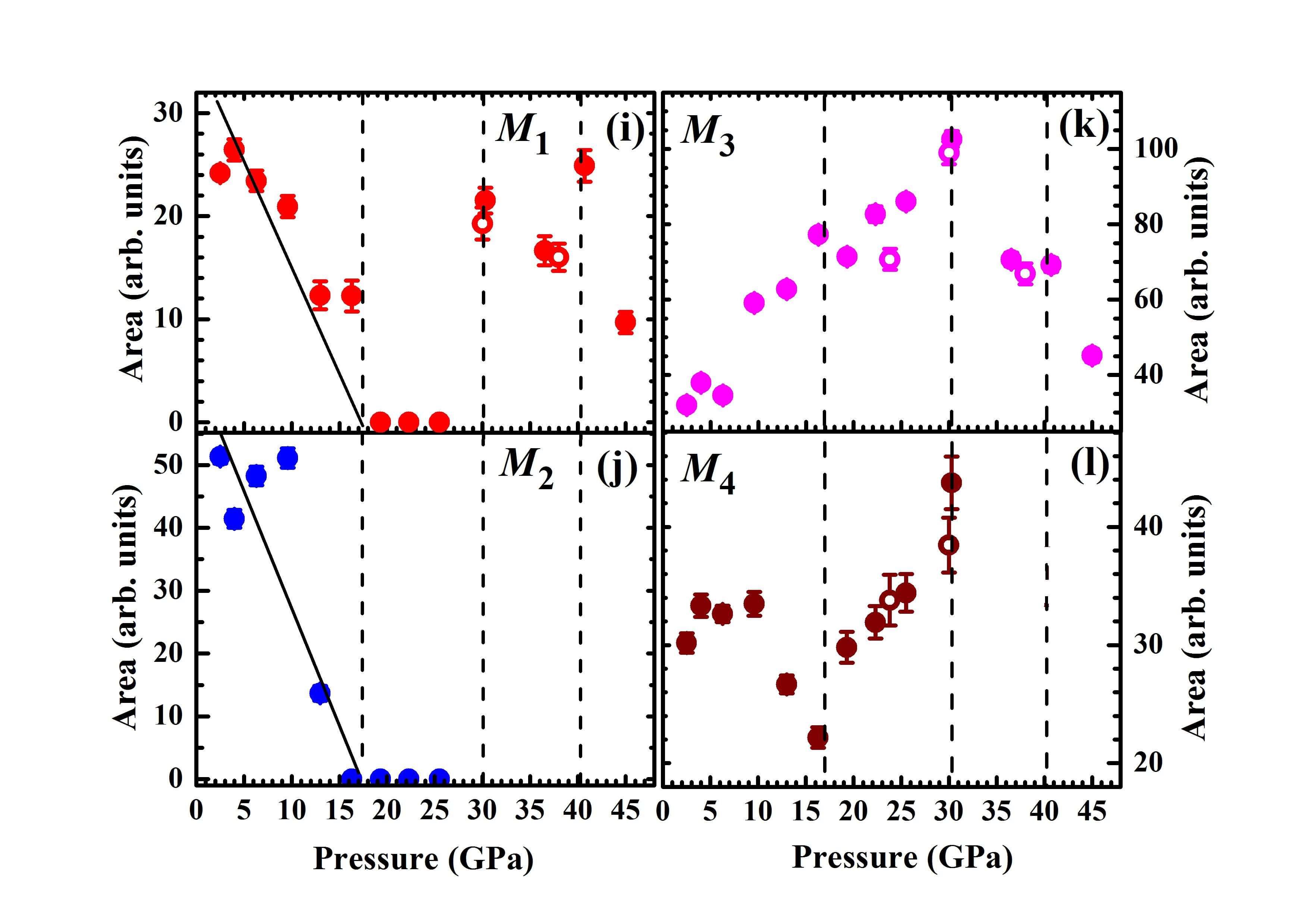}
	\begin{quotation}
		\caption{Pressure dependence of frequency (a-d), full width at half maximum (FWHM, e-h) and integrated area (i-l) of phonon Raman peaks $M_1$-$M_4$ (symbols) for the preliminary run. The vertical dashed lines mark the pressures $P_1 = 17$ GPa, $P_2=30$ GPa and $P_3=40$ GPa where phonon anomalies are noticed. Closed and open symbols represent data taken under increasing and decreasing pressure, respectively. The solid lines in (a-d) represent scalings to the unit cell volume according to the Gr\"uneisen's law (see text), and those in (i) and (j) are guides to the eyes.}
		\label{fitresults_run1}
	\end{quotation}
\end{figure}

\begin{figure}
	\includegraphics[width=0.45\textwidth]{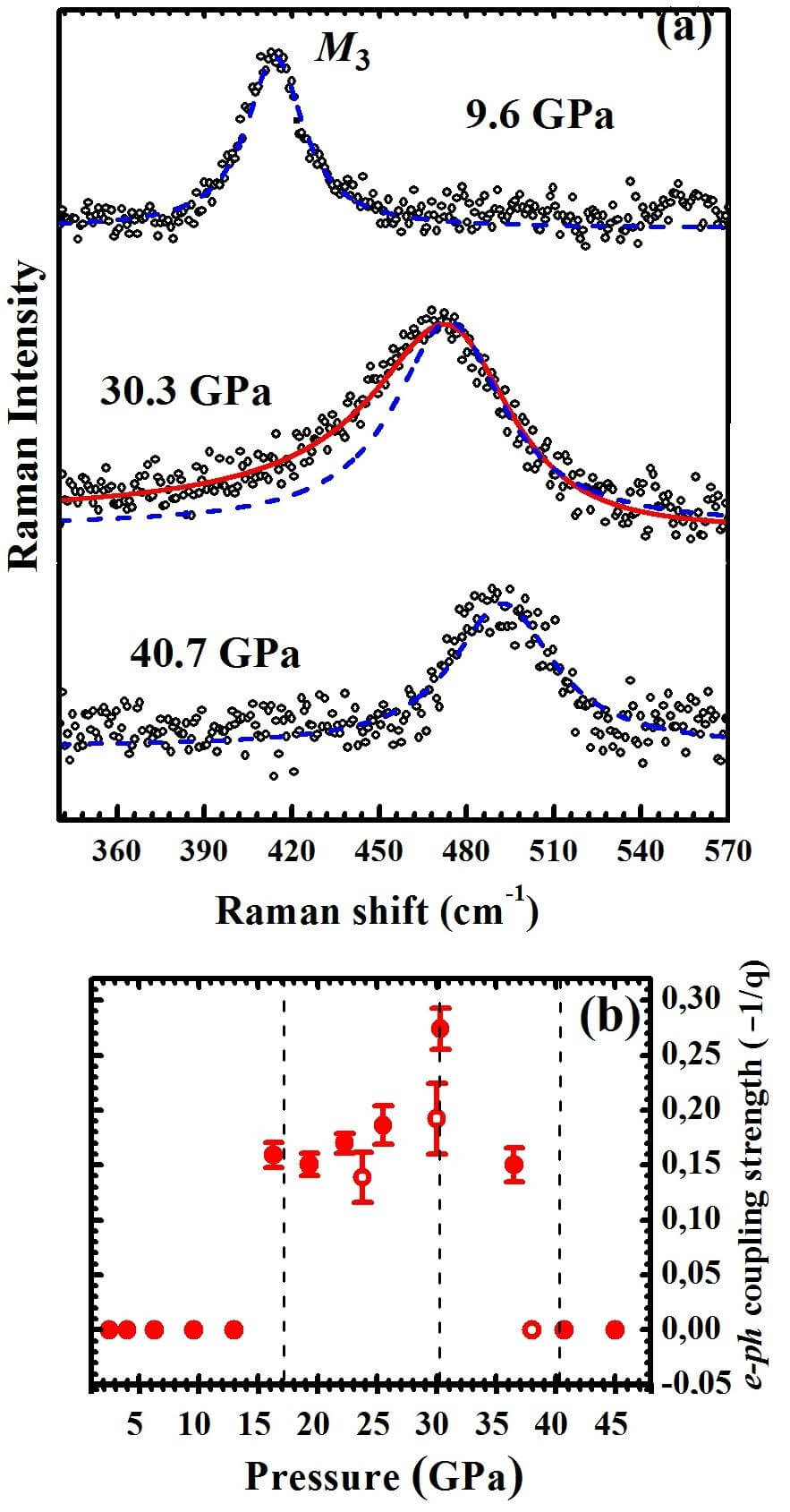}
	\begin{quotation}
		\caption{(a) Fittings of the $M_3$ mode profile at 9.6, 30.3 and 40.7 GPa; experimental data are represented by symbols; solid line represents the fitting using an asymmetric Fano lineshape at 30.3 GPa, while fits to symmetric Lorentz lineshapes at 9.6 and 40.7 GPa are represented by dashed lines. The Lorentz curve at 30.3 GPa is a guide to the eyes meant to highlight the peak asymmetric lineshape. (b) Electron-phonon (e-ph) coupling strength $\lvert 1/q \rvert$ obtained from the Fano lineshape fits as a function of pressure. Closed and open symbols represent data taken under increasing and releasing pressure, respectively.}
		\label{fano}
	\end{quotation}
\end{figure}

For the preliminary run, a relatively intense fluorescence signal from diamond impurities was superposed to the Raman signal. The baseline subtraction procedure to extract the phonon Raman peaks is illustrated in Fig. \ref{Raman_background}.
Figures \ref{fitresults_run1}(a)-\ref{fitresults_run1}(l) show the frequencies (a-d), linewidths (e-h) and areas (i-l) of modes $M_1-M_4$. Figure \ref{fano}(a) shows in detail the lineshape of mode $M_3$ at selected pressures, highlighting the asymmetric profile at intermediate pressures. The pressure-dependence of the Fano asymmetry parameter $\lvert 1/q \rvert$ is given in Fig. \ref{fano}(b). Overall, these preliminary data are consistent with the data shown in the main paper, arguing for the reproducibility of our results. A notable exception is the area of mode $M_4$ [see Fig. \ref{fitresults}(l) of the main paper and Fig. \ref{fitresults_run1}(l)], which we ascribe to a dificulty to extract the correct baseline nearby this mode in the preliminary data [see Fig. \ref{Raman_background}(a)]. 

\end{appendix}

\begin{acknowledgments}
We thank D. S. Rigitano, M. A. Eleot\'erio, and J. Fonseca J\'unior for experimental support. LNLS is acknowledged for beamtime concession. This work was supported by Fapesp Grants 2012/04870-7 and 2016/00756-6, and by CAPES and CNPq, Brazil.
\end{acknowledgments}

\newpage

\end{document}